\documentstyle{mn}

\newif\ifAMStwofonts

\ifoldfss
  \ifCUPmtlplainloaded \else
    \NewTextAlphabet{textbfit} {cmbxti10} {}
    \NewTextAlphabet{textbfss} {cmssbx10} {}
    \NewMathAlphabet{mathbfit} {cmbxti10} {} 
    \NewMathAlphabet{mathbfss} {cmssbx10} {} 
  \fi
  \ifAMStwofonts
    \ifCUPmtlplainloaded \else
      \NewSymbolFont{upmath} {eurm10}
      \NewSymbolFont{AMSa} {msam10}
      \NewMathSymbol{\upi}     {0}{upmath}{19}
      \NewMathSymbol{\umu}     {0}{upmath}{16}
      \NewMathSymbol{\upartial}{0}{upmath}{40}
      \NewMathSymbol{\leqslant}{3}{AMSa}{36}
      \NewMathSymbol{\geqslant}{3}{AMSa}{3E}

      \let\leq=\leqslant \let\le=\leqslant
      \let\geq=\geqslant \let\ge=\geqslant
    \fi
  \fi
\fi 

\ifnfssone
  \newmathalphabet{\mathit}
  \addtoversion{normal}{\mathit}{cmr}{m}{it}
  \addtoversion{bold}{\mathit}{cmr}{bx}{it}
  \newmathalphabet{\mathbfit} 
  \addtoversion{normal}{\mathbfit}{cmr}{bx}{it}
  \addtoversion{bold}{\mathbfit}{cmr}{bx}{it}
  \newmathalphabet{\mathbfss} 
  \addtoversion{normal}{\mathbfss}{cmss}{bx}{n}
  \addtoversion{bold}{\mathbfss}{cmss}{bx}{n}
  \ifAMStwofonts
    \ifCUPmtlplainloaded \else
      \UseAMStwoboldmath
      \makeatletter
      \new@mathgroup\upmath@group
      \define@mathgroup\mv@normal\upmath@group{eur}{m}{n}
      \define@mathgroup\mv@bold\upmath@group{eur}{b}{n}
      \edef\UPM{\hexnumber\upmath@group}
      \new@mathgroup\amsa@group
      \define@mathgroup\mv@normal\amsa@group{msa}{m}{n}
      \define@mathgroup\mv@bold\amsa@group{msa}{m}{n}
      \edef\AMSa{\hexnumber\amsa@group}
      \makeatother
      \mathchardef\upi="0\UPM19
      \mathchardef\umu="0\UPM16
      \mathchardef\upartial="0\UPM40
      \mathchardef\leqslant="3\AMSa36
      \mathchardef\geqslant="3\AMSa3E

      \let\leq=\leqslant \let\le=\leqslant
      \let\geq=\geqslant \let\ge=\geqslant
    \fi
  \fi
\fi 

\ifnfsstwo
  \DeclareMathAlphabet{\mathbfit}{OT1}{cmr}{bx}{it}
  \SetMathAlphabet\mathbfit{bold}{OT1}{cmr}{bx}{it}
  \DeclareMathAlphabet{\mathbfss}{OT1}{cmss}{bx}{n}
  \SetMathAlphabet\mathbfss{bold}{OT1}{cmss}{bx}{n}
  \ifAMStwofonts
    \ifCUPmtlplainloaded \else
      \DeclareSymbolFont{UPM}{U}{eur}{m}{n}
      \SetSymbolFont{UPM}{bold}{U}{eur}{b}{n}
      \DeclareSymbolFont{AMSa}{U}{msa}{m}{n}
      \DeclareMathSymbol{\upi}{0}{UPM}{"19}
      \DeclareMathSymbol{\umu}{0}{UPM}{"16}
      \DeclareMathSymbol{\upartial}{0}{UPM}{"40}
      \DeclareMathSymbol{\leqslant}{3}{AMSa}{"36}
      \DeclareMathSymbol{\geqslant}{3}{AMSa}{"3E}

      \let\leq=\leqslant \let\le=\leqslant
      \let\geq=\geqslant \let\ge=\geqslant
    \fi
  \fi
\fi 

\ifCUPmtlplainloaded \else
  \ifAMStwofonts \else 
    \def\upi{\pi}
    \def\umu{\mu}
    \def\upartial{\partial}
  \fi
\fi

\title{The CLASS blazar survey. I - selection criteria and radio properties}
\author[M. J. March\~a et al.]
       {M. J. March\~a,$^1$ \thanks{E-mail: mmarcha@oal.ul.pt} A. Caccianiga,$^1$ 
        I.W.A. Browne$^2$, N. Jackson,$^2$ \\
        $^1$CAAUL, Observat\'orio Astron\'omico de Lisboa, Tapada da Ajuda, 
1349-018 Lisboa, Portugal \\
        $^2$ University of Manchester, Jodrell Bank Observatory, Macclesfield, 
        Cheshire SK11 9DL, UK \\
        }
\date{}

\pagerange{\pageref{firstpage}--\pageref{lastpage}}
\pubyear{1994}

\begin{document}

\input{psfig.tex}
\maketitle

\label{firstpage}

\begin{abstract}

This paper presents a new complete and well defined sample of Flat
Spectrum Radio Sources (FSRS) selected from the CLASS survey with the
further constraint of a bright (mag$\leq$ 17.5) optical counterpart.
The sample has been designed to produce a large number of low-luminosity
blazars in order to test the current unifying models in the low
luminosity regime. In this first paper the new sample is presented and
the radio properties of the 325 sources contained therein discussed.

\end{abstract}


\section{Introduction}

The term {\it blazar} groups together two types of Active Galactic
Nuclei (AGNs): the Flat Spectrum Radio Quasars (FSRQs) and the BL Lac
objects.  The former, which are thought to be the beamed cores of FR~II
radio galaxies (Urry \& Padovani, 1995), are predominantly high
luminosity sources which show strong and broad emission features in
their spectra whereas the latter, usually associated with beamed cores
of FR~I radio galaxies (Urry \& Padovani, 1995), are primarily low
luminosity objects showing only weak or even absent emission features in
their spectra. Despite this dichotomy in the type of optical spectra,
and luminosity range, the 2 types of sources share a smooth non-thermal
continuum extending from radio to the X-rays, as well as the same
violent behavior such as the high degree of polarisation and
variability.

Recent years have seen a proliferation of efforts in order to find a
possible connection between the 2 types of blazars. The fact
that no basic differences were observed between the ratios of core to
extended radio luminosity and of X-rays to extended radio luminosity for BL
Lacs and FSRQs, led Maraschi \& Rovetti (1994) to propose that the 2
types of sources were basically similar, varying only in their
global intrinsic power. Later, Ghisellini (1997) and Fossati et al.
(1998) proposed that the intrinsic power was indeed the driving
feature behind the uniting sequence, but this time with the consequence
of shifting the peak of the Spectral Energy Distribution (SED). Indeed, as far
as the synchrotron component of the spectrum is concerned there seems to
be a sequence in the frequency at which the peak of the emission occurs:
low luminosity sources (i.e. the weakest BL Lacs) peak at high
frequencies (X-ray), while the high luminosity objects (i.e. FSRQ) peak
at low frequencies (infrared/sub-mm).  The position of the peak
frequency is thought to be related to the cooling rate of the electrons
responsible for the emission (Ghisellini 1997). 

With the aim of investigating the relationship between BL Lacs and other
Flat Spectrum Radio Sources (FSRS), March\~{a} et al. (1996) selected a
sample of low-luminosity, core-dominated radio sources with flux density
greater than 200~mJy at 5~GHz. By selecting sources with flat ($\alpha <
0.5$, S$_{\nu}\propto\nu^{-\alpha}$) radio spectra between 1.4 and 5~GHz
and bright ($V < 17$) counterparts, this 200~mJy sample was expected to
yield a purely radio flux limited sample out to $z = 0.1$. The objective was to
obtain a statistically well defined sample of low luminosity FSRS
consisting predominantly of optically passive sources, ie, sources with
weak emission features which could be identified with the beamed
counterparts of FRI radio galaxies.  Spectroscopic observations,
however, found an optical diversity which is difficult to understand
under the current unified schemes for radio-loud AGN in the low
luminosity regime. Apart from the expected BL Lacs and galaxies with
weak emission lines (PEGs for 'Passive Elliptical Galaxies'), the data
also yielded an unexpectedly high percentage ($\sim$20\%) of strong
emission-line objects. Among these, 2 sources were of particular
interest since they showed an unusual combination of BL Lac and quasar
properties. This 'hybrid' behavior is demonstrated by their low
luminosity, flat radio spectrum, a highly polarised optical continuum
(Jackson \& March\~a 1999), and broad emission lines.

Evidence for a sequence in the broadband properties of blazars
is found through the multi-wavelength study of the sources in the
200~mJy sample. Making use of data from across the spectrum, Ant\'on
(2000) finds a correlation between optical type, and the frequency at
which the synchrotron component peaks for the FSRS of the 200~mJy
sample. Specifically, BL Lacs tend to have the highest peak frequency,
and PEGs the lowest, whereas the 2 low luminosity blazars mentioned
before seem to have intermediate peak frequencies.

The 200~mJy sample, however, is relatively small (containing $\sim 60$
objects) which implies that the number of sources falling within each
category gives poor statistics. Hence, confirmation of these results
based on a more extensive study is a goal worth pursuing.  The ideal
sample should not only be larger than the 200~mJy, but also extend
itself to lower radio luminosities since the goal is to select a
sizeable number of low-power, low-redshift blazars. A search through the
literature reveals a few samples of FSRS, but none with the desired
specifications.  For instance, both the 1~Jy sample (K\"uhr et al. 1981)
containing 518 sources with flux densities at 5~GHz $S_{5GHz} \geq$ 1 Jy
and $V \leq 20$, and the CJ-F sample (Taylor et al. 1996) containing 293
sources with $S_{4.85GHz} \geq 350$ mJy and without specified magnitude
limit are samples of too high a flux density limit to be of interest for
the purpose in mind. The only other sample of FSRS available in the
literature is the DXRBS (Perlman et al. 1998) containing sources
resulting from a cross correlation between the WGACAT (White, Giommi, \&
Angelini 1994) X-ray catalogue, and a number of radio catalogues
publicly available. The sample contains 298 sources of which 106 new
identifications just appeared in the literature (Landt et al. 2001).
However, this sample is not specifically targeted at selecting
low-power, low-redshift BL Lacs, something that constitutes one of the
goals of the present work.

Given the current situation with the available samples of FSRS, it was
decided that the cleanest approach would be to select a new sample
following the criteria used for the 200~mJy, but this time making use of
a deeper survey. The starting point was the CLASS survey (Cosmic Lens
All Sky Survey, Myers et al. 2001).  The resulting sample goes to a
limiting flux density of 30 ~mJy at 5~GHz, i.e., roughly 6 times lower
than the flux density limit of the 200~mJy sample. In this paper the
selection criteria of this new sample are presented and its radio
properties discussed, leaving the discussion of the optical properties
to a forthcoming paper (Caccianiga et al.; submitted to MNRAS).

The paper is organised as follows. In the next section the radio survey
from which the sample was selected is briefly discussed. This is
followed by the definition of the selection criteria of the new sample.
Finally the radio properties of the sample are discussed and the summary
presented.

\section{The CLASS survey}

The CLASS survey is one of the deepest survey of flat spectrum radio
sources available so far. The primary goal of this survey was the
selection, in the radio band, of gravitational lenses to be used to
estimate the cosmological parameters.

The CLASS statistical sample, which contains $\sim 11,600$ sources, has
been defined by using the NVSS catalogue (Condon et al. 1998) at 1.4~GHz
and the GB6 catalogue (Gregory et al. 1996) at 5GHz.  These two surveys
offer the advantage of a large and uniform sky coverage at relatively
deep flux limits (about 2.5 mJy for the NVSS survey and about 25 mJy for
the GB6).  The CLASS catalogue is the result of a positional
cross-correlation between the two catalogues with the further constraint
of a flat $\alpha_{1.4}^{4.8}\leq$0.5, 
spectral index.

Since the two catalogues have been produced with instruments having
different resolutions (the NVSS beam has FWHM=45$\arcsec$ while the GB6
beam has FWHM$\sim$3.5$\arcmin$), some sources are resolved in
more components in the NVSS than in the GB6 catalogue where they appear
as a unique entry. As a consequence, the 1.4~GHz flux will be
systematically underestimated in these cases and the corresponding
spectral index will be flatter. This effect will bring some unwanted
steep spectrum sources in the sample.  In order to reduce this problem,
the positional cross-correlation has been done in such a way that all
the NVSS flux within a radius of 70$\arcsec$ of the GB6 position was
considered (Myers et al., in preparation).

However, this procedure will not eliminate completely the steep spectrum
sources from the sample because the Green Bank beam is uncapable of
resolving sources on scales up to $\sim$3-4$\arcmin$. Indeed, some of
these large double or triple radio sources have been found in the
catalogue (see section 4.1).

\section{The optically bright subsample}

The main goal of the work presented here is to select a sample of
low-luminosity blazars.  Given the typical absolute magnitudes of the
radio galaxies (Browne \& March\~a 1993) it is expected that all blazars
with z$\leq$0.15 will have R$\leq$17.5 due to the presence of the host
galaxy. Thus, the {\it optically bright sample} selected from the CLASS
survey by imposing the optical limit R $<$17.5 will contain as a subset
a purely radio flux limited sample for redshift below 0.15.

The selection of this sample was done by using the Automated Plate
Measuring machine facility (APM; {\it www.ast.cam.ac.uk/$\sim$apmcat/}).
Since APM covers only high galactic latitudes
($|b^{II}|\geq15-20^\circ$), the only objects considered were those with
$|b^{II}|\geq20^\circ$, where the catalogue is virtually complete.

Although APM offers the unique possibility of an automatic search for
the optical counterparts of entire catalogues containing a large number of
sources, there are two main problems which have to be addressed when
dealing with the automatic identification of optical counterparts:

\begin{enumerate}

\item Uncertainty on the optical position for optically bright
  (R$_{APM}\leq$15) and extended objects.

\item Uncertainty on the APM magnitude for extended objects.

\item APM blending or fragmentation of sources.

\end{enumerate}

\begin{figure}
\psfig{figure=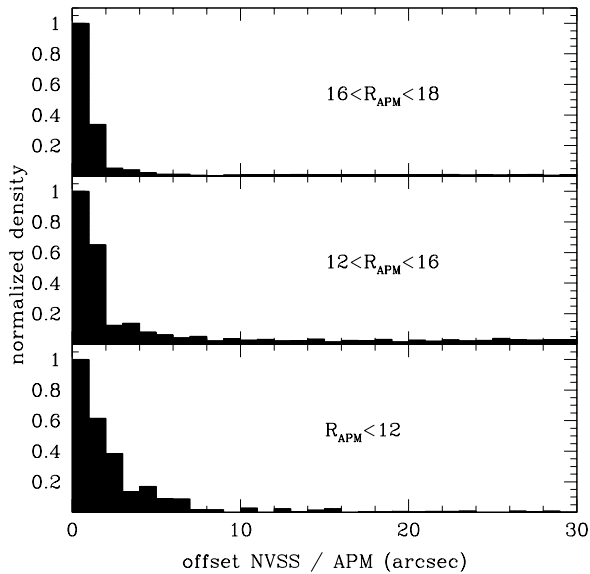,height=8cm,width=8cm}

 \caption{The normalized distribution of the offset between the CLASS and
the APM positions in three different ranges of magnitudes. The 
plotted quantity is the number of sources in each  bin divided
by the bin area.}
\label{offset}
\end{figure}

In order to address the first problem, the distribution of the NVSS/APM
positional offset has been studied in three different bins of magnitude:
R$_{APM}<$12, 12$\leq$ R$_{APM}<$16 and 16$\leq$ R$_{APM}<$18
(Fig.~\ref{offset}).  Positional offsets up to 30$\arcsec$ have been
considered.  The radius ($b$) of the circle that includes 95\% of the
optical counterparts has been computed by assuming that all the
radio/optical matches with an offset larger than $r_{max}$ are spurious.
For the bins 12$\leq$R$_{APM}<$16 and 16$\leq$R$_{APM}<$18 an
$r_{max}$=15$\arcsec$ was assumed, whereas for the brightest bin
(R$_{APM}<$12) an $r_{max}$=20$\arcsec$ was considered.

The resulting values of $b$ are: b=4$\arcsec$ (for
16$\leq$ R$_{APM}<$18), b=6$\arcsec$ (12$\leq$ R$_{APM}<$16) and
b=15$\arcsec$ (for R$_{APM}<$12).

The final list of sources was derived by keeping only the sources that
have a radio/optical offset below ${\it b}$ in the three magnitude
intervals.  In the case of the brightest bin (R$_{APM}\leq$12), where
the large positional tolerance can introduce some spurious radio/optical
matches, the optical finding charts were visually inspected and the
clear cases of wrong optical counterpart excluded.

The second issue to address in the optical identification process is the
correction of the APM magnitudes.  Due to the way the detection
algorithm works, APM magnitudes for sources which are extended and
brighter than $\sim$17 can be systematically underestimated (Ant\'on
2000). In particular, the APM red magnitudes (brighter than 15-16)
appear systematically underestimated when compared with photometric
magnitudes in R band (which is close to the band used for the POSS
plates). This discrepancy appears more evident for extended objects
(galaxies).  However, besides this systematic discrepancy, a large
dispersion (up to 2 magnitudes) is also observed. It was thus decided
that an empirical correction should be applied to the red magnitude, in
order to reduce these two problems.  This correction has the following
form (Ant\'on 2000):

\begin{center}

$R = 7.235 + 0.605 R_{APM}+0.013\sigma_R$

\end{center}

where $R$ is the corrected R magnitude, $R_{APM}$ is the observed APM
red magnitude  and $\sigma_R$ gives the departure from a Gaussian
profile in units of $\sigma$ (given in the APM output). The
magnitude shown in Table~1 is this corrected magnitude.

Finally the problem of APM blending and fragmentation of sources was
resolved by visual inspection of all of the finding charts. In this way
it was possible to manually correct the few cases where the magnitude of
a source was completely miscalculated due to the blending of 2 nearby
objects.

In summary, the optically bright CLASS sample contains 325 sources which
obey the following selection criteria:

\begin{enumerate}
\item {$35\degr \leq\delta\leq75\degr$}
\item {$|b^{II}|\geq$20$^\circ$}
\item {$S_{5}\geq$ 30 mJy}
\item {flat spectrum, i.e. $\alpha_{1.4}^{4.8}\leq$0.5
(S$_{\nu}\propto\nu^{-\alpha}$)}
\item {APM red magnitude (corrected) $\leq$ 17.5}
\end{enumerate}

The complete sample of optically bright CLASS sources is presented in
Table~1. The columns are: (1) - name, based on the GB6 position; (2) -
the NVSS position (J2000); (3) - APM corrected red magnitude; (4) - the
NVSS flux density (1.4~GHz). This is the sum of the fluxes of all the
NVSS sources found within 70$\arcsec$ from the GB6 position (in only 38
cases more than one NVSS component has been found); (5) - the GB6 flux
density (4.8~GHz); (6) - VLA flux density at 8.4~GHz (``$<$'' indicates
upper limit for the detection); (7) - the spectral index between 1.4
and 4.8~GHz; (8) - the spectral index between 1.4 and 8.4~GHz (a ``$>$''
indicates an lower limit based on the detection limit in column 6); (9) - a
flag indicating whether the source has been re-classified as steep,
between 1.4 and 4.8~GHz on the basis of a re-analysis of the NVSS map.
The corrected values of $\alpha_{1.4}^{4.8}$ are reported in parenthesis 
(see the next section for details).

\setcounter{table}{0}
\begin{table*}
 \centering
 \begin{minipage}{140mm}
\label{sample}
\caption{The CLASS bright sample  (continued on the next page)}
\begin{tabular}{c c r r r r r r c}
name & Position (J2000) & R mag & S$_{1.4}$ & S$_{4.8}$ & S$_{8.4}$ 
& $\alpha_{1.4}^{4.8}$ & $\alpha_{1.4}^{8.4}$ & flag \\ 
 & & & (mJy) & (mJy) & (mJy) & & & \\
 & & & & & & & & \\

GB6J001450+385512 & 00 14 52.82 +38 55 19.3 & 17.00 &         48 &    47 &         28 &  0.02 &  0.30 &   \\
GB6J002538+400830 & 00 25 38.19 +40 08 42.3 & 15.23 &         37 &    32 &         19 &  0.12 &  0.37 &   \\
GB6J005529+354443 & 00 55 30.56 +35 44 35.4 & 17.00 &         49 &    32 &         27 &  0.35 &  0.33 &   \\
GB6J011216+381910 & 01 12 18.07 +38 18 56.9 & 17.28 &         58 &    52 &        126 &  0.09 & -0.43 &   \\
GB6J011920+385253 & 01 19 20.13 +38 53 07.4 & 15.27 &         99 &    69 &         23 &  0.29 &  0.81 & s (0.63)\\
GB6J012626+395406 & 01 26 25.82 +39 54 13.1 & 16.31 &        175 &    95 &       $<$2 &  0.50 & $>$2.50  &   \\
GB6J013631+390623 & 01 36 32.43 +39 05 59.4 & 16.38 &         61 &    49 &         42 &  0.17 &  0.20 &   \\
GB6J014156+392325 & 01 41 57.67 +39 23 29.1 & 14.42 &        117 &    80 &         69 &  0.31 &  0.29 &   \\
GB6J015451+362746 & 01 54 51.59 +36 27 45.9 & 15.28 &        234 &   154 &         69 &  0.34 &  0.68 &   \\
GB6J020405+400512 & 02 04 05.16 +40 05 03.4 & 15.79 &         49 &   103 &        136 & -0.60 & -0.57 &   \\
GB6J021625+400101 & 02 16 24.55 +40 00 35.9 & 14.39 &         42 &    49 &         36 & -0.12 &  0.09 &   \\
GB6J022526+371029 & 02 25 27.29 +37 10 27.9 & 14.01 &        197 &   137 &        167 &  0.29 &  0.09 &   \\
GB6J024621+353634 & 02 46 21.05 +35 36 38.0 & 17.50 &        190 &   108 &        128 &  0.46 &  0.22 &   \\
GB6J025359+362539 & 02 54 00.03 +36 25 51.5 & 12.39 &         83 &    63 &         53 &  0.22 &  0.25 &   \\
GB6J055253+724045 & 05 52 53.00 +72 40 45.5 & 17.37 &        508 &   385 &        256 &  0.23 &  0.38 &   \\
GB6J061641+663024 & 06 16 42.13 +66 30 24.1 & 13.43 &        288 &   174 &        126 &  0.41 &  0.46 &   \\
GB6J064204+675834 & 06 42 04.23 +67 58 35.6 & 17.03 &        193 &   488 &        419 & -0.75 & -0.43 &   \\
GB6J065010+605001 & 06 50 08.63 +60 50 44.7 & 12.38 &         63 &    44 &         25 &  0.30 &  0.52 &   \\
GB6J065422+504223 & 06 54 22.13 +50 42 23.5 & 16.13 &        198 &   136 &        305 &  0.30 & -0.24 &   \\
GB6J065648+560258 & 06 56 47.58 +56 03 09.2 & 16.00 &         56 &    67 &         59 & -0.14 & -0.03 &   \\
GB6J065817+501641 & 06 58 18.49 +50 16 14.9 & 16.00 &         47 &    30 &          7 &  0.36 &  1.06 & s (0.90) \\
GB6J070105+693634 & 07 01 06.53 +69 36 29.5 & 17.36 &        484 &   314 &        252 &  0.35 &  0.36 &   \\
GB6J070506+421457 & 07 05 07.17 +42 14 50.1 & 14.50 &         47 &    32 &          4 &  0.32 &  1.38 &   \\
GB6J070648+592327 & 07 06 49.54 +59 23 14.7 & 16.31 &         51 &    49 &         34 &  0.04 &  0.23 &   \\
GB6J070904+483659 & 07 09 07.83 +48 36 53.2 & 12.57 &        376 &   259 &         86 &  0.30 &  0.82 & s (0.87)\\
GB6J070932+501056 & 07 09 34.27 +50 10 56.4 & 12.30 &        121 &   126 &        167 & -0.03 & -0.18 &   \\
GB6J071044+422053 & 07 10 44.30 +42 20 54.7 & 17.47 &        335 &   236 &        216 &  0.28 &  0.25 &   \\
GB6J071045+473203 & 07 10 46.16 +47 32 11.2 & 15.10 &       1019 &   984 &        620 &  0.03 &  0.28 &   \\
GB6J071433+740811 & 07 14 35.93 +74 08 09.7 & 17.30 &        116 &   216 &        134 & -0.50 & -0.08 &   \\
GB6J071510+452554 & 07 15 10.01 +45 25 55.5 & 14.51 &         76 &    98 &         90 & -0.21 & -0.09 &   \\
GB6J071547+413926 & 07 15 42.94 +41 39 23.5 & 16.15 &        104 &    56 &         12 &  0.50 &  1.20 &   \\
GB6J072028+370647 & 07 20 28.77 +37 06 45.2 & 16.95 &         13 &    36 &         37 & -0.86 & -0.61 &   \\
GB6J072151+712036 & 07 21 53.10 +71 20 36.7 & 15.96 &        727 &   859 &        594 & -0.14 &  0.11 &   \\
GB6J072403+535121 & 07 24 05.73 +53 51 10.0 & 17.50 &         18 &    30 &         18 & -0.43 & -0.01 &   \\
GB6J072849+570124 & 07 28 49.59 +57 01 22.8 & 16.67 &        488 &   387 &        644 &  0.19 & -0.15 &   \\
GB6J073328+351532 & 07 33 29.58 +35 15 42.4 & 16.80 &        102 &    70 &         34 &  0.31 &  0.61 &   \\
GB6J073502+475011 & 07 35 02.25 +47 50 07.9 & 17.45 &        405 &   511 &        460 & -0.19 & -0.07 &   \\
GB6J073654+653603 & 07 36 48.44 +65 36 41.5 &  8.93 &         57 &    51 &       $<$2 &  0.09 & $>$1.87 & s (0.82)\\
GB6J073728+594106 & 07 37 30.08 +59 41 03.4 & 13.69 &        561 &   370 &        248 &  0.34 &  0.46 &   \\
GB6J073758+643048 & 07 37 58.98 +64 30 43.4 & 17.13 &        417 &   251 &        239 &  0.41 &  0.31 &   \\
GB6J073933+495449 & 07 39 34.91 +49 54 39.0 & 16.11 &        101 &    55 &         48 &  0.49 &  0.42 &   \\
GB6J074904+451027 & 07 49 06.41 +45 10 33.8 & 17.14 &        159 &   113 &         43 &  0.28 &  0.73 &   \\
GB6J075038+654114 & 07 50 34.51 +65 41 23.4 & 16.95 &         32 &    52 &         27 & -0.39 &  0.10 & s (0.68)\\
GB6J080036+501047 & 08 00 36.03 +50 10 44.1 & 17.32 &        111 &    63 &         38 &  0.46 &  0.60 &   \\
GB6J080053+392433 & 08 00 53.56 +39 24 40.8 & 13.31 &         56 &    47 &         4  &  0.13 &  1.41 & s (0.97)\\
GB6J080132+473624 & 08 01 31.88 +47 36 16.9 & 16.64 &         77 &    65 &         59 &  0.14 &  0.15 & s (0.76)\\
GB6J080346+655424 & 08 03 47.78 +65 54 33.0 & 17.40 &         75 &    55 &         40 &  0.25 &  0.35 &   \\
GB6J080624+593059 & 08 06 25.94 +59 31 06.8 & 17.28 &         61 &    38 &         61 &  0.38 &  0.00 &   \\
GB6J080839+495033 & 08 08 39.74 +49 50 36.1 & 17.50 &       1115 &  1315 &        880 & -0.13 &  0.13 &   \\
GB6J080949+521856 & 08 09 49.23 +52 18 58.1 & 15.69 &        183 &   184 &        154 & -0.01 &  0.10 &   \\
GB6J081245+650843 & 08 12 41.08 +65 09 11.2 & 17.20 &         31 &    35 &         37 & -0.10 & -0.10 &   \\
GB6J081622+573858 & 08 16 22.68 +57 39 09.2 & 17.44 &        100 &    64 &         92 &  0.36 &  0.05 &   \\
GB6J082154+503136 & 08 21 53.77 +50 31 19.3 & 17.32 &         59 &    53 &         33 &  0.09 &  0.33 &   \\
GB6J082437+405712 & 08 24 38.88 +40 57 07.9 & 17.36 &        178 &   138 &         93 &  0.21 &  0.36 &   \\
GB6J083055+540041 & 08 31 00.36 +54 00 24.4 & 15.27 &         17 &    40 &         41 & -0.71 & -0.50 &   \\
GB6J083140+460800 & 08 31 39.80 +46 08 00.2 & 16.17 &        132 &    98 &         87 &  0.24 &  0.23 &   \\
GB6J083411+580318 & 08 34 11.06 +58 03 21.4 & 15.72 &         58 &    57 &         33 &  0.01 &  0.31 &   \\
GB6J083455+553430 & 08 34 54.91 +55 34 21.0 & 17.18 &       8284 &  5740 &       3234 &  0.30 &  0.52 &   \\
GB6J083901+401608 & 08 39 03.22 +40 15 46.6 & 16.81 &         46 &    34 &         19 &  0.25 &  0.50 &   \\
GB6J084124+705345 & 08 41 24.46 +70 53 41.4 & 17.31 &       3824 &  2342 &       1746 &  0.40 &  0.44 &   \\
GB6J084215+452547 & 08 42 15.43 +45 25 43.0 & 17.23 &         75 &    54 &         53 &  0.26 &  0.19 &   \\
GB6J084241+391100 & 08 42 41.78 +39 10 53.1 & 16.47 &         56 &    33 &       $<$2 &  0.42 & $>$1.86 & \\

\end{tabular}
\end{minipage}

\end{table*}

\begin{table*}
\contcaption{}
 \centering
 \begin{minipage}{140mm}
\begin{tabular}{c c r r r r r r c}
name & Position (J2000) & R mag & S$_{1.4}$ & S$_{4.8}$ & S$_{8.4}$ 
& $\alpha_{1.4}^{4.8}$ & $\alpha_{1.4}^{8.4}$ & flag \\ 
 & & & (mJy) & (mJy) & (mJy) & & & \\
 & & & & & & & & \\

GB6J084352+374232 & 08 43 52.68 +37 42 27.0 & 17.10 &        104 &    61 &         45 &  0.43 &  0.47 &   \\
GB6J084356+510522 & 08 43 59.09 +51 05 25.5 & 16.16 &        201 &   116 &         44 &  0.45 &  0.85 & s (0.57)\\
GB6J085005+403610 & 08 50 04.76 +40 36 07.4 & 17.29 &        133 &   110 &        144 &  0.15 & -0.05 &   \\
GB6J085008+593054 & 08 50 10.00 +59 31 17.6 & 16.54 &         19 &    37 &         32 & -0.52 & -0.28 &   \\
GB6J085317+682824 & 08 53 19.15 +68 28 17.3 & 12.07 &        293 &   183 &         58 &  0.38 &  0.90 &   \\
GB6J085416+532731 & 08 54 17.56 +53 27 35.6 & 17.32 &         25 &    32 &         15 & -0.20 &  0.29 &   \\
GB6J085449+621900 & 08 54 50.51 +62 18 50.5 & 17.44 &        388 &   215 &        317 &  0.48 &  0.11 &   \\
GB6J090536+470603 & 09 05 36.56 +47 05 46.5 & 17.10 &        102 &    86 &         63 &  0.14 &  0.27 &   \\
GB6J090615+463633 & 09 06 15.58 +46 36 19.0 & 16.65 &        291 &   159 &        206 &  0.49 &  0.19 &   \\
GB6J090643+710018 & 09 06 43.70 +71 00 30.2 & 16.46 &         55 &    31 &         15 &  0.46 &  0.72 &   \\
GB6J090650+412426 & 09 06 52.78 +41 24 29.1 & 14.52 &         54 &    65 &         81 & -0.16 & -0.23 &   \\
GB6J090757+493558 & 09 07 56.36 +49 35 48.3 & 15.00 &         43 &    36 &         28 &  0.14 &  0.24 &   \\
GB6J092230+710926 & 09 22 30.09 +71 09 36.3 & 17.28 &         83 &    83 &         56 &  0.00 &  0.22 &   \\
GB6J092914+501323 & 09 29 15.53 +50 13 35.5 & 17.30 &        523 &   544 &        747 & -0.03 & -0.20 &   \\
GB6J092932+625637 & 09 29 35.49 +62 56 59.1 & 15.34 &         56 &    33 &         15 &  0.42 &  0.73 &   \\
GB6J093254+673654 & 09 32 53.91 +67 37 07.0 & 12.16 &        238 &   131 &         28 &  0.48 &  1.19 &   \\
GB6J093737+723106 & 09 37 31.93 +72 30 55.2 & 16.44 &         26 &    39 &         35 & -0.34 & -0.17 &   \\
GB6J094319+361447 & 09 43 19.16 +36 14 52.2 & 12.94 &         75 &    81 &        335 & -0.06 & -0.83 &   \\
GB6J094542+575739 & 09 45 42.29 +57 57 46.2 & 16.87 &        125 &    92 &         69 &  0.25 &  0.33 &   \\
GB6J094557+461907 & 09 45 57.17 +46 19 18.6 & 16.74 &         27 &    31 &         57 & -0.11 & -0.42 &   \\
GB6J094614+581932 & 09 46 14.92 +58 19 34.5 & 16.23 &         87 &    56 &         12 &  0.35 &  1.10 &   \\
GB6J094832+553538 & 09 48 32.03 +55 35 35.6 & 15.78 &         31 &    34 &         22 & -0.07 &  0.19 &   \\
GB6J095227+504837 & 09 52 27.15 +50 48 50.2 & 17.44 &         89 &   104 &        189 & -0.12 & -0.42 &   \\
GB6J095301+720231 & 09 52 58.29 +72 02 41.2 & 16.42 &        110 &    62 &         13 &  0.46 &  1.19 &   \\
GB6J095531+690357 & 09 55 33.28 +69 03 55.0 &  7.89 &         89 &    98 &         97 & -0.08 & -0.05 &   \\
GB6J095552+694047 & 09 55 51.88 +69 40 46.1 &  9.00 &       6437 &  3796 &         49 &  0.43 &  2.72 &   \\
GB6J095736+552258 & 09 57 38.18 +55 22 57.4 & 17.21 &       3080 &  2015 &       1498 &  0.34 &  0.40 &   \\
GB6J095847+653405 & 09 58 47.22 +65 33 54.3 & 16.58 &        730 &  1125 &       1224 & -0.35 & -0.29 &   \\
GB6J100055+533158 & 10 00 54.58 +53 32 05.5 & 17.39 &        109 &    72 &         28 &  0.34 &  0.76 &   \\
GB6J100308+681313 & 10 03 06.65 +68 13 17.4 & 16.25 &        103 &    86 &         64 &  0.15 &  0.27 &   \\
GB6J100712+502346 & 10 07 10.42 +50 23 55.8 & 16.53 &         31 &    34 &         24 & -0.08 &  0.14 &   \\
GB6J100724+580201 & 10 07 24.89 +58 02 03.8 & 17.50 &        186 &   135 &         85 &  0.26 &  0.44 &   \\
GB6J101028+413230 & 10 10 27.63 +41 32 36.6 & 16.88 &        499 &   854 &        449 & -0.44 &  0.06 &   \\
GB6J101244+423009 & 10 12 44.22 +42 29 56.7 & 17.40 &         80 &    46 &         41 &  0.44 &  0.37 &   \\
GB6J101504+492606 & 10 15 04.13 +49 26 01.1 & 16.67 &        378 &   299 &        252 &  0.19 &  0.23 &   \\
GB6J101859+591126 & 10 18 58.61 +59 11 27.9 & 17.26 &         98 &    76 &         61 &  0.21 &  0.26 &   \\
GB6J102310+394759 & 10 23 11.60 +39 48 17.2 & 17.32 &       1123 &   789 &        779 &  0.29 &  0.20 &   \\
GB6J102521+372641 & 10 25 23.42 +37 26 50.3 & 14.46 &         64 &    35 &          7 &  0.49 &  1.24 &   \\
GB6J102632+750501 & 10 26 30.53 +75 05 17.1 & 17.50 &        113 &    83 &         42 &  0.25 &  0.55 &   \\
GB6J103053+411300 & 10 30 53.56 +41 13 15.0 & 14.42 &         51 &    51 &         54 &  0.00 & -0.03 & s? (0.65)\\
GB6J103118+505350 & 10 31 18.51 +50 53 36.9 & 16.76 &         38 &    34 &         32 &  0.09 &  0.09 &   \\
GB6J103123+744158 & 10 31 22.57 +74 41 58.2 & 17.00 &        218 &   250 &        103 & -0.11 &  0.42 &   \\
GB6J103318+422228 & 10 33 18.23 +42 22 37.2 & 17.34 &         46 &    34 &         16 &  0.24 &  0.58 &   \\
GB6J103550+375646 & 10 35 51.25 +37 56 41.3 & 17.44 &         53 &    34 &         54 &  0.36 & -0.01 &   \\
GB6J103653+444832 & 10 36 52.95 +44 48 18.6 & 17.31 &         42 &    66 &         70 & -0.37 & -0.29 &   \\
GB6J103742+571158 & 10 37 44.28 +57 11 56.4 & 17.37 &         72 &   126 &         82 & -0.46 & -0.07 &   \\
GB6J103951+405557 & 10 39 51.04 +40 55 41.8 & 16.17 &         29 &    48 &         36 & -0.42 & -0.13 &   \\
GB6J104630+544953 & 10 46 28.73 +54 49 45.3 & 16.85 &         82 &    77 &         43 &  0.05 &  0.36 &   \\
GB6J104910+373742 & 10 49 09.82 +37 37 59.7 & 17.20 &         53 &    30 &         15 &  0.47 &  0.71 &   \\
GB6J105115+464439 & 10 51 15.97 +46 44 17.7 & 17.48 &        321 &   270 &        200 &  0.14 &  0.26 &   \\
GB6J105203+424203 & 10 52 04.15 +42 41 53.3 & 16.96 &         97 &    70 &         48 &  0.26 &  0.39 &   \\
GB6J105344+493006 & 10 53 44.02 +49 29 55.2 & 15.72 &         65 &    59 &         40 &  0.07 &  0.27 &   \\
GB6J105349+402337 & 10 53 49.14 +40 23 52.0 & 15.31 &         73 &    42 &          8 &  0.45 &  1.24 & s (0.72)\\
GB6J105430+385500 & 10 54 31.80 +38 55 21.6 & 17.25 &         62 &    56 &         17 &  0.08 &  0.72 &   \\
GB6J105730+405631 & 10 57 31.12 +40 56 46.5 & 11.71 &         45 &    31 &         14 &  0.30 &  0.65 &   \\
GB6J105837+562817 & 10 58 37.71 +56 28 11.8 & 16.11 &        229 &   247 &        189 & -0.06 &  0.11 &   \\
GB6J110147+722536 & 11 01 48.52 +72 25 37.4 & 17.42 &       1246 &   858 &        366 &  0.30 &  0.68 &   \\
GB6J110242+594132 & 11 02 42.65 +59 41 19.7 & 17.37 &        352 &   249 &        177 &  0.28 &  0.38 &   \\
GB6J110428+381228 & 11 04 27.42 +38 12 32.9 & 13.30 &        816 &   723 &        631 &  0.10 &  0.14 &   \\
GB6J110508+465311 & 11 05 07.05 +46 53 18.9 & 17.15 &         60 &    44 &         35 &  0.25 &  0.30 &   \\
GB6J110552+394649 & 11 05 53.69 +39 46 55.1 & 17.00 &         45 &    53 &         44 & -0.13 &  0.01 &   \\
GB6J110657+603345 & 11 06 56.76 +60 34 02.4 & 16.54 &         33 &    31 &         28 &  0.04 &  0.09 &   \\

\end{tabular}
\end{minipage}
\end{table*}

\begin{table*}
 \centering
 \begin{minipage}{140mm}
\contcaption{}
\begin{tabular}{c c r r r r r r c}
name & Position (J2000) & R mag & S$_{1.4}$ & S$_{4.8}$ & S$_{8.4}$ 
& $\alpha_{1.4}^{4.8}$ & $\alpha_{1.4}^{8.4}$ & flag \\ 
 & & & (mJy) & (mJy) & (mJy) & & & \\
 & & & & & & & & \\

GB6J110939+383046 & 11 09 39.19 +38 31 21.5 & 17.00 &         19 &    58 &         79 & -0.93 & -0.81 &   \\
GB6J111106+522751 & 11 11 05.75 +52 27 49.6 & 17.35 &        116 &    74 &         36 &  0.36 &  0.65 &   \\
GB6J111206+352707 & 11 12 08.06 +35 27 06.8 & 12.99 &         64 &    49 &        121 &  0.22 & -0.36 &   \\
GB6J111309+401827 & 11 13 05.93 +40 17 32.8 & 15.00 &         13 &    45 &         19 & -0.99 & -0.20 & s? (1.2) \\
GB6J111903+602832 & 11 19 04.13 +60 28 43.3 & 16.01 &         60 &    33 &       $<$2 &  0.49 & $>$1.90 &   \\
GB6J111912+623938 & 11 19 16.56 +62 39 26.6 & 15.50 &         47 &    43 &         45 &  0.06 &  0.02 &   \\
GB6J111914+600459 & 11 19 13.90 +60 04 55.9 & 17.31 &        264 &   170 &        126 &  0.36 &  0.41 &   \\
GB6J112047+421206 & 11 20 48.06 +42 12 14.3 & 17.28 &         24 &    30 &         27 & -0.18 & -0.06 &   \\
GB6J112157+431459 & 11 21 56.66 +43 14 58.8 & 17.29 &         49 &    37 &         23 &  0.22 &  0.42 &   \\
GB6J112413+513350 & 11 24 13.14 +51 33 50.1 & 17.01 &         55 &    39 &         31 &  0.28 &  0.32 &   \\
GB6J112832+583322 & 11 28 32.61 +58 33 46.8 & 11.80 &        678 &   392 &         30 &  0.44 &  1.74 &   \\
GB6J113626+700931 & 11 36 26.48 +70 09 25.8 & 13.91 &        340 &   267 &        213 &  0.20 &  0.26 &   \\
GB6J113629+673707 & 11 36 29.92 +67 37 06.0 & 16.65 &         46 &    47 &         37 & -0.02 &  0.12 &   \\
GB6J114047+462207 & 11 40 47.91 +46 22 04.0 & 16.02 &         99 &    54 &         22 &  0.49 &  0.84 &   \\
GB6J114115+595309 & 11 41 16.00 +59 53 09.2 & 14.24 &        119 &   147 &        106 & -0.17 &  0.06 &   \\
GB6J114300+730413 & 11 43 04.73 +73 04 09.3 & 16.90 &         49 &    32 &          6 &  0.34 &  1.17 &   \\
GB6J114312+612214 & 11 43 12.10 +61 22 11.1 & 17.49 &         84 &    94 &         99 & -0.09 & -0.09 &   \\
GB6J114722+350109 & 11 47 22.14 +35 01 07.3 & 16.00 &        638 &   669 &        499 & -0.04 &  0.14 &   \\
GB6J114850+592459 & 11 48 50.42 +59 24 57.3 & 14.15 &        482 &   566 &        521 & -0.13 & -0.04 &   \\
GB6J114856+525432 & 11 48 56.63 +52 54 25.7 & 16.74 &         93 &   288 &        597 & -0.91 & -1.04 &   \\
GB6J114959+552832 & 11 50 00.15 +55 28 21.8 & 16.82 &        153 &    83 &         70 &  0.50 &  0.44 &   \\
GB6J115037+653916 & 11 50 34.48 +65 39 28.8 & 17.08 &         78 &    60 &         49 &  0.21 &  0.26 &   \\
GB6J115126+585913 & 11 51 24.66 +58 59 18.6 & 17.41 &        185 &   131 &        102 &  0.28 &  0.33 &   \\
GB6J115757+552713 & 11 57 56.14 +55 27 12.6 & 11.47 &        101 &    88 &        157 &  0.12 & -0.24 &   \\
GB6J120209+444452 & 12 02 08.43 +44 44 20.8 & 17.47 &        106 &    69 &         54 &  0.35 &  0.38 &   \\
GB6J120304+603130 & 12 03 03.55 +60 31 19.4 & 16.00 &        191 &   182 &        191 &  0.04 &  0.00 &   \\
GB6J120328+480316 & 12 03 29.92 +48 03 13.7 & 17.17 &         63 &   164 &        415 & -0.77 & -1.05 &   \\
GB6J120334+451050 & 12 03 35.49 +45 10 49.8 & 17.46 &         39 &    43 &         89 & -0.08 & -0.46 &   \\
GB6J120856+464112 & 12 08 55.57 +46 41 13.7 & 16.16 &         65 &    62 &         32 &  0.04 &  0.39 & s (0.86)\\
GB6J120922+411938 & 12 09 22.82 +41 19 41.1 & 17.12 &        274 &   459 &        485 & -0.42 & -0.32 &   \\
GB6J121008+355224 & 12 10 08.05 +35 52 42.4 & 15.41 &         26 &    44 &         39 & -0.44 & -0.23 &   \\
GB6J121108+503000 & 12 11 11.37 +50 29 44.0 & 12.15 &          5 &    60 &         15 & -1.97 & -0.58 &   \\
GB6J121331+504446 & 12 13 29.28 +50 44 30.3 & 12.51 &         97 &    86 &         61 &  0.10 &  0.26 &   \\
GB6J121510+462710 & 12 15 09.99 +46 27 15.9 & 17.13 &        272 &   168 &        229 &  0.39 &  0.10 &   \\
GB6J121541+361924 & 12 15 39.98 +36 19 26.5 & 15.32 &         35 &    32 &       $<$2 &  0.07 & $>$1.59 &   \\
GB6J121558+354313 & 12 15 59.38 +35 43 02.5 & 17.27 &         55 &    30 &          1 &  0.48 &  2.04 &   \\
GB6J121736+515502 & 12 17 36.72 +51 55 10.8 & 17.04 &         85 &    61 &         33 &  0.27 &  0.53 &   \\
GB6J122208+581427 & 12 22 09.40 +58 14 21.5 & 16.89 &         52 &    51 &         35 &  0.02 &  0.22 &   \\
GB6J122306+582659 & 12 23 02.04 +58 26 40.7 & 11.97 &        134 &    81 &         27 &  0.41 &  0.89 &   \\
GB6J122405+500130 & 12 24 09.92 +50 01 56.7 & 17.39 &         47 &    31 &         20 &  0.34 &  0.48 &   \\
GB6J123012+470031 & 12 30 11.80 +47 00 23.0 & 13.04 &         94 &    73 &         51 &  0.21 &  0.34 &   \\
GB6J123132+641421 & 12 31 31.34 +64 14 19.4 & 16.47 &         59 &    36 &         44 &  0.40 &  0.16 &   \\
GB6J123151+353929 & 12 31 51.76 +35 39 59.3 & 16.58 &         39 &    32 &         40 &  0.16 & -0.01 &   \\
GB6J123350+502630 & 12 33 49.24 +50 26 23.4 & 17.44 &        283 &   175 &         64 &  0.39 &  0.83 &   \\
GB6J123413+475408 & 12 34 13.37 +47 53 51.6 & 17.21 &        356 &   268 &        176 &  0.23 &  0.39 &   \\
GB6J123417+505441 & 12 34 16.31 +50 54 25.7 & 16.94 &         66 &    48 &         41 &  0.26 &  0.26 &   \\
GB6J123532+522839 & 12 35 30.60 +52 28 28.1 & 17.44 &         88 &    80 &         68 &  0.08 &  0.14 &   \\
GB6J124036+695837 & 12 40 34.64 +69 58 31.2 & 17.50 &        133 &   190 &        164 & -0.29 & -0.12 &   \\
GB6J124307+731549 & 12 43 10.31 +73 16 00.9 & 14.20 &        493 &   312 &        153 &  0.37 &  0.65 &   \\
GB6J124313+362755 & 12 43 12.70 +36 27 45.1 & 17.41 &        148 &    91 &         53 &  0.39 &  0.57 &   \\
GB6J124732+672322 & 12 47 33.31 +67 23 16.8 & 16.65 &        264 &   174 &        131 &  0.34 &  0.39 &   \\
GB6J124818+582029 & 12 48 18.77 +58 20 28.8 & 15.80 &        245 &   356 &        324 & -0.30 & -0.16 &   \\
GB6J125311+530113 & 12 53 11.94 +53 01 12.1 & 17.13 &        488 &   363 &        366 &  0.24 &  0.16 &   \\
GB6J125614+565220 & 12 56 14.13 +56 52 23.8 & 12.81 &        318 &   415 &        256 & -0.22 &  0.12 &   \\
GB6J130132+463357 & 13 01 32.61 +46 34 03.4 & 16.68 &         87 &   155 &        166 & -0.47 & -0.36 &   \\
GB6J130146+441612 & 13 01 46.35 +44 16 19.9 & 17.36 &         58 &    47 &         48 &  0.18 &  0.11 &   \\
GB6J130836+434405 & 13 08 37.90 +43 44 15.8 & 12.59 &         59 &    47 &         35 &  0.18 &  0.29 &   \\
GB6J130924+430502 & 13 09 25.58 +43 05 05.5 & 16.97 &         60 &    45 &         31 &  0.23 &  0.37 &   \\
GB6J131215+445023 & 13 12 16.81 +44 50 21.0 & 12.23 &        140 &   106 &         39 &  0.23 &  0.71 &   \\
GB6J131218+351522 & 13 12 17.75 +35 15 21.5 & 16.71 &         48 &    36 &         28 &  0.23 &  0.30 &   \\
GB6J131328+363538 & 13 13 27.28 +36 35 40.4 & 09.00 &        123 &    79 &          3 &  0.36 &  1.99 &   \\
GB6J131655+722619 & 13 16 58.95 +72 26 20.8 & 17.47 &         40 &    76 &         61 & -0.51 & -0.23 &   \\
GB6J131739+411538 & 13 17 39.18 +41 15 46.4 & 13.42 &        267 &   195 &        171 &  0.25 &  0.25 &   \\

\end{tabular}
\end{minipage}
\end{table*}

\begin{table*}
 \centering
 \begin{minipage}{140mm}
\contcaption{}
\begin{tabular}{c c r r r r r r c}
name & Position (J2000) & R mag & S$_{1.4}$ & S$_{4.8}$ & S$_{8.4}$ 
& $\alpha_{1.4}^{4.8}$ & $\alpha_{1.4}^{8.4}$ & flag \\ 
 & & & (mJy) & (mJy) & (mJy) & & & \\
 & & & & & & & & \\

GB6J131947+514759 & 13 19 46.40 +51 48 06.7 & 17.39 &       1093 &   614 &        234 &  0.47 &  0.86 &   \\
GB6J132513+395610 & 13 25 13.34 +39 55 53.7 & 14.46 &         58 &    53 &         27 &  0.07 &  0.43 &   \\
GB6J134035+444801 & 13 40 35.25 +44 48 17.7 & 16.73 &         78 &    73 &         51 &  0.06 &  0.24 &   \\
GB6J134139+371653 & 13 41 38.81 +37 16 45.3 & 17.14 &        123 &    77 &         28 &  0.38 &  0.82 &   \\
GB6J134442+402811 & 13 44 41.55 +40 28 02.3 & 13.38 &        102 &    79 &          7 &  0.21 &  1.50 & s (0.77)\\
GB6J134444+555322 & 13 44 42.21 +55 53 13.1 & 13.52 &        145 &    95 &         37 &  0.35 &  0.76 &   \\
GB6J134856+395904 & 13 48 55.95 +39 59 07.3 & 12.12 &         83 &    62 &         67 &  0.24 &  0.12 &   \\
GB6J134913+601114 & 13 49 15.29 +60 11 26.5 & 14.54 &         79 &    43 &         12 &  0.50 &  1.05 &   \\
GB6J134934+534125 & 13 49 34.72 +53 41 17.2 & 17.40 &       1097 &   644 &        764 &  0.43 &  0.20 &   \\
GB6J135251+654124 & 13 52 51.23 +65 41 14.1 & 17.10 &        129 &    70 &         79 &  0.50 &  0.28 &   \\
GB6J135313+350912 & 13 53 14.28 +35 08 48.3 & 16.26 &         43 &    31 &         28 &  0.27 &  0.24 &   \\
GB6J135327+401700 & 13 53 26.67 +40 16 58.8 & 10.91 &         41 &    35 &         32 &  0.13 &  0.14 &   \\
GB6J135607+413637 & 13 56 07.37 +41 36 14.9 & 16.90 &         27 &    40 &         38 & -0.33 & -0.20 &   \\
GB6J140850+650550 & 14 08 48.77 +65 05 28.3 & 17.44 &         35 &    43 &         48 & -0.18 & -0.18 &   \\
GB6J141132+742404 & 14 11 34.73 +74 24 29.4 & 17.35 &        107 &    82 &         73 &  0.21 &  0.21 &   \\
GB6J141159+423952 & 14 11 59.89 +42 39 48.9 & 17.15 &         92 &    50 &         45 &  0.49 &  0.40 &   \\
GB6J141343+433959 & 14 13 43.68 +43 39 45.5 & 16.06 &         49 &    39 &         27 &  0.18 &  0.33 &   \\
GB6J141536+483102 & 14 15 36.77 +48 30 30.6 & 17.43 &         42 &    37 &         39 &  0.10 &  0.04 &   \\
GB6J141946+542328 & 14 19 46.50 +54 23 15.1 & 15.53 &        814 &  1350 &       2247 & -0.41 & -0.57 &   \\
GB6J142312+505543 & 14 23 14.17 +50 55 38.7 & 16.65 &        309 &   232 &        214 &  0.23 &  0.21 &   \\
GB6J142814+391222 & 14 28 13.84 +39 12 18.4 & 17.47 &        252 &   169 &        138 &  0.32 &  0.34 &   \\
GB6J143120+395245 & 14 31 20.50 +39 52 41.8 & 16.73 &        219 &   259 &        235 & -0.14 & -0.04 &   \\
GB6J143239+361823 & 14 32 39.82 +36 18 08.2 & 13.50 &        119 &   137 &         88 & -0.11 &  0.17 &   \\
GB6J143646+633645 & 14 36 45.68 +63 36 37.5 & 17.40 &        952 &   757 &        867 &  0.19 &  0.05 &   \\
GB6J143920+371148 & 14 39 20.52 +37 12 03.4 & 17.49 &         60 &    77 &         57 & -0.21 &  0.02 &   \\
GB6J144920+422103 & 14 49 20.73 +42 21 01.3 & 17.23 &        156 &   118 &         96 &  0.23 &  0.27 &   \\
GB6J150411+685605 & 15 04 12.51 +68 56 14.8 & 17.18 &        133 &   227 &        112 & -0.43 &  0.10 & s? (0.55)\\
GB6J150522+604007 & 15 05 22.67 +60 40 13.3 & 15.69 &         56 &    35 &          3 &  0.39 &  1.64 &   \\
GB6J150854+373318 & 15 08 54.94 +37 33 28.3 & 15.34 &         15 &    39 &       $<$2 & -0.79 & $>$1.11 &   \\
GB6J150914+700436 & 15 09 13.02 +70 04 22.9 & 14.81 &         65 &    47 &         13 &  0.27 &  0.90 &   \\
GB6J151554+561850 & 15 15 56.27 +56 18 53.2 & 11.12 &         54 &    31 &       $<$2 &  0.44 & $>$1.83 & s (1.15)\\
GB6J151717+694715 & 15 17 14.61 +69 47 10.2 & 17.50 &         15 &    32 &         21 & -0.64 & -0.21 &   \\
GB6J151746+652456 & 15 17 47.55 +65 25 23.6 & 17.43 &         38 &    31 &         36 &  0.16 &  0.03 &   \\
GB6J151806+424346 & 15 18 06.20 +42 44 42.5 & 13.33 &         51 &    32 &          7 &  0.38 &  1.11 &   \\
GB6J151807+665746 & 15 18 08.95 +66 57 53.4 & 12.92 &         39 &    36 &         30 &  0.06 &  0.14 &   \\
GB6J151838+404532 & 15 18 38.93 +40 45 00.7 & 15.45 &         44 &    44 &         27 &  0.01 &  0.28 &   \\
GB6J152911+693455 & 15 29 14.41 +69 34 59.3 & 17.28 &         54 &    70 &         34 & -0.22 &  0.26 &   \\
GB6J153134+720634 & 15 31 33.54 +72 06 41.4 & 17.50 &        419 &   444 &        232 & -0.05 &  0.33 &   \\
GB6J153900+353053 & 15 39 01.66 +35 30 46.1 & 16.50 &         97 &    92 &         79 &  0.04 &  0.11 &   \\
GB6J153931+381011 & 15 39 32.11 +38 09 52.5 & 15.61 &         52 &    40 &         25 &  0.21 &  0.40 &   \\
GB6J154255+612950 & 15 42 56.94 +61 29 55.5 & 17.25 &         87 &   121 &        144 & -0.27 & -0.28 &   \\
GB6J154255+705000 & 15 42 55.44 +70 49 41.7 & 12.85 &         46 &    47 &         38 & -0.02 &  0.10 &   \\
GB6J154504+525930 & 15 45 04.84 +52 59 26.0 & 17.48 &        208 &   122 &         64 &  0.43 &  0.66 &   \\
GB6J155158+580642 & 15 51 58.18 +58 06 44.8 & 16.70 &        191 &   348 &        305 & -0.49 & -0.26 &   \\
GB6J155722+544043 & 15 57 21.47 +54 40 20.0 & 13.64 &        114 &    85 &         48 &  0.24 &  0.48 & s? (0.53)\\
GB6J155848+562524 & 15 58 48.30 +56 25 14.4 & 17.31 &        208 &   206 &        169 &  0.01 &  0.12 &   \\
GB6J155901+592437 & 15 59 01.67 +59 24 21.5 & 12.88 &        220 &   191 &        130 &  0.11 &  0.29 &   \\
GB6J155922+731026 & 15 59 28.95 +73 10 27.2 & 16.79 &         46 &    32 &       $<$2 &  0.29 & $>$1.74 &   \\
GB6J160029+613743 & 16 00 29.53 +61 37 26.7 & 13.94 &        104 &    61 &       $<$2 &  0.43 & $>$2.20 & s (0.65)\\
GB6J160318+694552 & 16 03 18.49 +69 45 57.4 & 17.12 &        203 &   172 &        102 &  0.13 &  0.38 &   \\
GB6J160357+573101 & 16 03 55.95 +57 30 54.8 & 17.39 &        332 &   365 &        308 & -0.08 &  0.04 &   \\
GB6J160532+531250 & 16 05 32.23 +53 12 59.5 & 14.66 &         63 &    43 &         26 &  0.31 &  0.49 &   \\
GB6J160820+601834 & 16 08 20.67 +60 18 29.5 & 16.67 &         62 &    59 &         35 &  0.04 &  0.32 &   \\
GB6J161447+374554 & 16 14 46.93 +37 46 07.4 & 16.68 &         50 &    36 &         67 &  0.26 & -0.16 &   \\
GB6J161941+525617 & 16 19 42.44 +52 56 13.5 & 17.38 &        183 &   128 &        181 &  0.29 &  0.00 &   \\
GB6J161947+523319 & 16 19 45.99 +52 33 13.3 & 14.59 &         49 &    30 &       $<$2 &  0.40 & $>$1.78 &   \\
GB6J162025+690512 & 16 20 26.29 +69 04 47.6 & 17.16 &         72 &    46 &         25 &  0.36 &  0.59 &   \\
GB6J162303+662411 & 16 23 04.44 +66 24 01.0 & 17.00 &        156 &   481 &        302 & -0.91 & -0.37 &   \\

\end{tabular}
\end{minipage}
\end{table*}

\begin{table*}
 \centering
 \begin{minipage}{140mm}
\contcaption{}
\begin{tabular}{c c r r r r r r c}
name & Position (J2000) & R mag & S$_{1.4}$ & S$_{4.8}$ & S$_{8.4}$ 
& $\alpha_{1.4}^{4.8}$ & $\alpha_{1.4}^{8.4}$ & flag \\ 
 & & & (mJy) & (mJy) & (mJy) & & & \\
 & & & & & & & & \\

GB6J162308+390946 & 16 23 07.60 +39 09 32.4 & 16.62 &        180 &   244 &        219 & -0.25 & -0.11 &   \\
GB6J162509+405345 & 16 25 10.35 +40 53 34.4 & 12.21 &        164 &   110 &         86 &  0.32 &  0.36 &   \\
GB6J162612+512044 & 16 26 11.61 +51 20 39.3 & 17.30 &         50 &    35 &         21 &  0.29 &  0.49 &   \\
GB6J162636+580914 & 16 26 37.42 +58 09 11.7 & 16.50 &        534 &   315 &        157 &  0.43 &  0.68 &   \\
GB6J163801+552547 & 16 38 00.70 +55 25 25.2 & 13.61 &        164 &   107 &          5 &  0.34 &  1.92 &   \\
GB6J163813+572029 & 16 38 13.41 +57 20 24.2 & 17.09 &       1199 &  1750 &       1326 & -0.31 & -0.06 &   \\
GB6J164220+665608 & 16 42 21.79 +66 55 49.3 & 17.29 &        127 &    73 &         47 &  0.45 &  0.55 &   \\
GB6J164258+394842 & 16 42 58.77 +39 48 37.0 & 16.12 &       7099 &  8719 &       5439 & -0.17 &  0.15 &   \\
GB6J164420+454644 & 16 44 20.05 +45 46 45.4 & 17.06 &        188 &   109 &         64 &  0.44 &  0.60 &   \\
GB6J164734+494954 & 16 47 35.13 +49 49 57.2 & 16.49 &        182 &   191 &        189 & -0.04 & -0.02 &   \\
GB6J165138+400227 & 16 51 37.64 +40 02 17.2 & 17.39 &         44 &    49 &         31 & -0.10 &  0.19 &   \\
GB6J165353+394541 & 16 53 52.24 +39 45 36.6 & 11.50 &       1562 &  1375 &       2772 &  0.10 & -0.32 &   \\
GB6J165414+434319 & 16 54 12.49 +43 43 10.6 & 17.50 &         85 &    46 &         16 &  0.50 &  0.93 &   \\
GB6J165503+540754 & 16 55 05.07 +54 07 57.2 & 16.50 &         50 &    31 &          7 &  0.38 &  1.09 &   \\
GB6J165547+444735 & 16 55 47.36 +44 47 25.0 & 15.15 &         38 &    35 &         37 &  0.07 &  0.02 &   \\
GB6J165721+570556 & 16 57 20.76 +57 05 54.5 & 17.40 &        940 &   764 &        521 &  0.17 &  0.33 &   \\
GB6J165728+741302 & 16 57 32.33 +74 12 51.3 & 16.99 &          8 &    51 &         62 & -1.50 & -1.14 &   \\
GB6J170123+395432 & 17 01 24.70 +39 54 36.2 & 17.27 &        191 &   150 &        285 &  0.19 & -0.22 &   \\
GB6J170449+713840 & 17 04 47.05 +71 38 16.9 & 16.81 &         36 &    43 &         30 & -0.15 &  0.10 &   \\
GB6J170716+453607 & 17 07 17.77 +45 36 11.7 & 17.45 &        795 &   461 &        319 &  0.44 &  0.51 &   \\
GB6J171523+572434 & 17 15 22.94 +57 24 40.5 & 11.26 &         58 &    35 &         26 &  0.40 &  0.44 &   \\
GB6J171613+683636 & 17 16 13.94 +68 36 38.3 & 17.37 &        489 &   838 &        819 & -0.44 & -0.29 &   \\
GB6J171718+422711 & 17 17 19.18 +42 26 59.6 & 16.00 &        134 &   125 &         81 &  0.06 &  0.28 &   \\
GB6J171813+422759 & 17 18 15.20 +42 28 19.2 & 17.24 &        112 &    86 &         62 &  0.21 &  0.33 &   \\
GB6J171914+485839 & 17 19 14.59 +48 58 49.3 & 13.76 &        146 &   164 &        206 & -0.10 & -0.19 &   \\
GB6J171937+480404 & 17 19 38.31 +48 04 12.2 & 16.20 &         64 &   109 &        188 & -0.44 & -0.60 &   \\
GB6J171941+354700 & 17 19 36.23 +35 47 06.0 & 17.41 &         34 &    30 &       $<$2 &  0.11 & $>$1.59 &   \\
GB6J172110+354217 & 17 21 09.53 +35 42 16.5 & 17.18 &        821 &   784 &        601 &  0.04 &  0.17 &   \\
GB6J172315+654751 & 17 23 14.13 +65 47 46.1 & 16.99 &        239 &   182 &        166 &  0.22 &  0.20 &   \\
GB6J172535+585127 & 17 25 35.07 +58 51 39.3 & 17.12 &         76 &    55 &         49 &  0.26 &  0.24 &   \\
GB6J172722+551059 & 17 27 23.49 +55 10 53.9 & 17.31 &        144 &   274 &        265 & -0.52 & -0.34 &   \\
GB6J172818+501315 & 17 28 18.58 +50 13 11.3 & 15.68 &        228 &   145 &        161 &  0.37 &  0.19 &   \\
GB6J172859+383819 & 17 28 59.20 +38 38 26.6 & 17.27 &        242 &   219 &        187 &  0.08 &  0.14 &   \\
GB6J173047+371451 & 17 30 46.88 +37 14 55.0 & 17.18 &        102 &    78 &         68 &  0.22 &  0.23 &   \\
GB6J173234+712359 & 17 32 32.75 +71 24 03.9 & 14.29 &        182 &   187 &       $<$2 & -0.02 & $>$2.52 & s (0.96)\\
GB6J173312+704632 & 17 33 12.89 +70 46 32.5 & 14.53 &         81 &    71 &          8 &  0.10 &  1.29 & s (1.00)\\
GB6J173410+421933 & 17 34 13.53 +42 19 57.3 & 15.26 &         51 &    36 &          8 &  0.28 &  1.03 &   \\
GB6J174113+722447 & 17 41 22.62 +72 24 51.5 & 17.46 &         89 &    52 &         63 &  0.43 &  0.19 &   \\
GB6J174231+594513 & 17 42 31.96 +59 45 07.3 & 17.08 &        107 &    88 &        116 &  0.16 & -0.05 &   \\
GB6J174455+554220 & 17 44 56.50 +55 42 16.6 & 12.39 &        707 &   562 &        305 &  0.19 &  0.47 &   \\
GB6J174832+700550 & 17 48 32.88 +70 05 51.6 & 16.93 &        736 &   715 &        572 &  0.02 &  0.14 &   \\
GB6J174900+432151 & 17 49 00.21 +43 21 51.8 & 17.49 &        281 &   321 &        284 & -0.11 & -0.01 &   \\
GB6J174914+595351 & 17 49 20.96 +59 53 10.9 & 15.00 &        116 &    63 &       $<$2 &  0.49 & $>$2.27 &   \\
GB6J175041+395706 & 17 50 41.17 +39 57 00.3 & 14.50 &         42 &    38 &         30 &  0.08 &  0.19 &   \\
GB6J175041+585132 & 17 50 40.58 +58 51 20.9 & 15.02 &         63 &    35 &          9 &  0.48 &  1.09 &   \\
GB6J175045+370858 & 17 50 44.75 +37 09 34.3 & 16.50 &         48 &    35 &         13 &  0.26 &  0.73 &   \\
GB6J175131+471259 & 17 51 31.62 +47 13 23.0 & 17.15 &         48 &    49 &         30 & -0.01 &  0.27 &   \\
GB6J175546+623652 & 17 55 48.36 +62 36 44.4 & 11.00 &        288 &   203 &        147 &  0.28 &  0.38 &   \\
GB6J175628+580708 & 17 56 29.19 +58 06 58.2 & 16.89 &         52 &    38 &         24 &  0.25 &  0.43 &   \\
GB6J175704+535153 & 17 57 06.74 +53 51 37.3 & 15.07 &         60 &    42 &         39 &  0.29 &  0.24 &   \\
GB6J175728+552309 & 17 57 28.37 +55 23 11.7 & 14.11 &         80 &    73 &         50 &  0.07 &  0.26 &   \\
GB6J175833+663801 & 17 58 33.41 +66 37 58.8 & 13.07 &        772 &   875 &       $<$2 & -0.10 & $>$3.32 &   \\
GB6J180132+440409 & 18 01 32.29 +44 04 20.6 & 17.23 &        727 &  1193 &        522 & -0.40 &  0.18 &   \\
GB6J180228+481932 & 18 02 31.66 +48 19 40.9 & 16.50 &         20 &    33 &         29 & -0.39 & -0.20 &   \\
GB6J180557+574749 & 18 05 57.64 +57 47 46.5 & 17.00 &         89 &    48 &         12 &  0.50 &  1.12 & s (0.66)\\
GB6J180651+694931 & 18 06 50.46 +69 49 28.1 & 15.06 &       2032 &  2122 &       1595 & -0.04 &  0.14 &   \\
GB6J180738+563159 & 18 07 37.75 +56 31 56.8 & 15.50 &         25 &    32 &         13 & -0.21 &  0.36 &   \\
GB6J181156+521431 & 18 11 57.12 +52 14 26.5 & 17.09 &         61 &    51 &         60 &  0.14 &  0.00 &   \\
GB6J181349+472123 & 18 13 50.19 +47 21 14.9 & 14.50 &        136 &    78 &         14 &  0.45 &  1.27 &   \\
GB6J181912+551103 & 18 19 10.12 +55 11 08.2 & 17.03 &         77 &    67 &         64 &  0.11 &  0.10 &   \\
GB6J181932+362230 & 18 19 32.87 +36 22 32.9 & 16.89 &        105 &    61 &         21 &  0.44 &  0.90 &   \\
GB6J183750+641842 & 18 37 50.06 +64 18 40.7 & 16.90 &         40 &    35 &         45 &  0.10 & -0.07 &   \\
GB6J183850+480237 & 18 38 49.22 +48 02 35.0 & 17.37 &         30 &    41 &         32 & -0.25 & -0.03 &   \\

\end{tabular}
\end{minipage}
\end{table*}

\begin{table*}
 \centering
 \begin{minipage}{140mm}
\contcaption{}
\begin{tabular}{c c r r r r r r c}
name & Position (J2000) & R mag & S$_{1.4}$ & S$_{4.8}$ & S$_{8.4}$ 
& $\alpha_{1.4}^{4.8}$ & $\alpha_{1.4}^{8.4}$ & flag \\ 
 & & & (mJy) & (mJy) & (mJy) & & & \\
 & & & & & & & & \\

GB6J183858+573535 & 18 38 58.66 +57 35 38.1 & 16.93 &         88 &    93 &         66 & -0.05 &  0.16 &   \\
GB6J184033+621257 & 18 40 33.53 +62 12 49.5 & 16.50 &         67 &    72 &         65 & -0.06 &  0.01 &   \\
GB6J184214+470952 & 18 42 15.26 +47 09 44.7 & 16.00 &         61 &    40 &         11 &  0.34 &  0.95 & s (0.78)\\
GB6J184917+670548 & 18 49 15.90 +67 05 40.9 & 16.47 &        518 &   845 &        476 & -0.40 &  0.05 &   \\
GB6J184941+642522 & 18 49 40.73 +64 25 15.4 & 15.00 &         74 &    47 &         13 &  0.37 &  0.97 &   \\
GB6J185328+504652 & 18 53 32.26 +50 46 12.3 & 16.50 &         34 &    31 &          6 &  0.07 &  0.97 & s? (0.99)\\
GB6J185455+735112 & 18 54 57.41 +73 51 20.7 & 16.89 &        465 &   576 &        600 & -0.17 & -0.14 &   \\
GB6J185852+682747 & 18 58 55.20 +68 27 39.8 & 15.77 &         52 &    30 &          2 &  0.44 &  1.74 &   \\
GB6J190042+645242 & 19 00 46.13 +64 52 57.4 & 16.83 &        105 &    59 &         10 &  0.47 &  1.31 &   \\
GB6J190101+620727 & 19 01 03.11 +62 07 21.2 & 16.87 &        109 &    68 &         23 &  0.38 &  0.87 &   \\
GB6J191212+660826 & 19 12 07.38 +66 07 46.9 & 15.11 &         67 &    39 &         21 &  0.44 &  0.65 &   \\
GB6J192325+740407 & 19 23 23.05 +74 04 05.1 & 15.00 &        195 &   111 &       $<$2 &  0.46 & $>$2.56 &   \\
GB6J192747+735755 & 19 27 48.53 +73 58 02.1 & 16.50 &       3951 &  3626 &       3697 &  0.07 &  0.04 &   \\
GB6J194553+705545 & 19 45 53.64 +70 55 48.3 & 17.12 &        953 &   677 &        482 &  0.28 &  0.38 &   \\
GB6J230115+351252 & 23 01 14.45 +35 13 00.6 & 16.83 &         90 &   129 &        109 & -0.29 & -0.11 &   \\
GB6J232349+365047 & 23 23 50.22 +36 51 04.2 & 14.96 &         71 &    68 &         24 &  0.03 &  0.60 &   \\
GB6J235352+385549 & 23 53 50.59 +38 55 54.9 & 17.08 &         48 &    31 &         20 &  0.36 &  0.49 &   \\

\end{tabular}
\end{minipage}
\end{table*}

\section{Radio properties}

The primary goal of this work is the selection of a sample of blazars,
i.e.  core-dominated, flat radio spectrum sources. However, the simple
cross-correlation between the NVSS and the GB6 catalogue and the
subsequent constraint on the radio spectral index can select some
objects which are not truly FSRS and/or core-dominated sources. Hence,
all additional available information to check the spectral indices and
the radio morphology of the sources contained in the sample was
considered. Specifically, the following pieces of information have been
used:

\begin{itemize}
\item {\bf VLA snap-shots at 8.4~GHz}. The entire CLASS catalogue
has been observed at 8.4~GHz with VLA in A-configuration. Except
for few (15) sources not detected at this frequency 
there is a flux density and a position at 8.4~GHz for 95\%
of the sources in the optically bright CLASS sample. 

\item {\bf FIRST data}. The FIRST survey (Becker et al. 1995, White et al.
1997) is made at 1.4~GHz with VLA in B configuration which is a good
compromise between resolution (beam$\sim$4$\arcsec$) and sensitivity
to extended emission (largest angular scale
$\theta_{LAS}\sim$100$\arcsec$). Hence, it is ideal to estimate the
compactness parameter for the CLASS sources. About 50\% of the sources
in the sample fall in the region of sky covered by the FIRST survey.

\end{itemize}

\subsection{The steep spectrum sources}

\begin{figure}
\psfig{figure=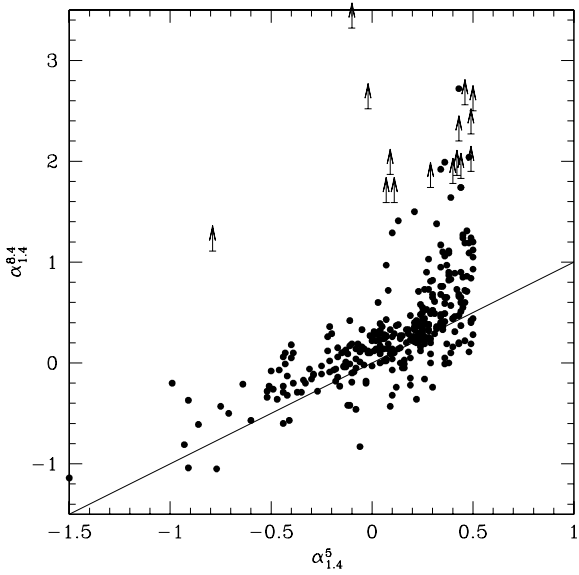,height=8cm,width=8cm}
 \caption{The spectral index between 1.4 and 8.4~GHz versus the spectral
   index between 1.4 and 4.8~GHz for the objects in the CLASS survey
   presented in this paper. The solid line represents the expected
   one-to-one relationship between the two indices; arrows correspond to
   lower limits based on the flux density
   detectability limits for the 8.4~GHz measurements.}
\label{slopes3}
\end{figure}

By definition, the CLASS sources have a flat spectrum between 1.4 and
4.8~GHz ($\alpha_{1.4}^{4.8}$, S$\sim\nu^{-\alpha}$ ) as computed from
the NVSS and the GB6 fluxes. Making use of the 8.4~GHz mentioned above,
a new spectral index was determined, this time between 1.4 and 8.4~GHz
($\alpha_{1.4}^{8.4}$). Figure~\ref{slopes2} shows the distribution of
$\alpha_{1.4}^{4.8}$ and $\alpha_{1.4}^{8.4}$ for the sample while
Figure~\ref{slopes3} shows $\alpha_{1.4}^{8.4}$ versus
$\alpha_{1.4}^{4.8}$.

In general, the two indices are well correlated although a dispersion is
clearly observed.  If the steepest objects are excluded (i.e., those
with $\alpha_{1.4}^{8.4} \geq$ 0.7), the slope of the correlation is
consistent with a one-to-one relationship.  The dispersion of the points
on this fixed one-to-one relationship is $\sigma_y$=0.25. Given this
dispersion, it is expected that a number of objects will have a rather
steep spectrum between 1.4 and 8.4~GHz, even if they were selected to
have a flat spectrum between 1.4 and 4.8~GHz. More quantitatively, about
10 objects are expected to have $\alpha_{1.4}^{8.4}\geq$0.75.  The
observed number, however, is much higher (60, including the lower
limits) thus indicating that there is a clear ``tail'' of steep objects
that cannot be explained simply as due to the statistical dispersion on
the $\alpha_{1.4}^{4.8}$/$\alpha_{1.4}^{8.4}$ correlation.  This
``tail'' is clearly visible also in Figure~\ref{slopes2}b.
 
The 60 sources with $\alpha_{1.4}^{8.4}\geq$0.75 were submitted to
further inspection of the NVSS maps and were gathered in 2 groups:

 \begin{figure}
\psfig{figure=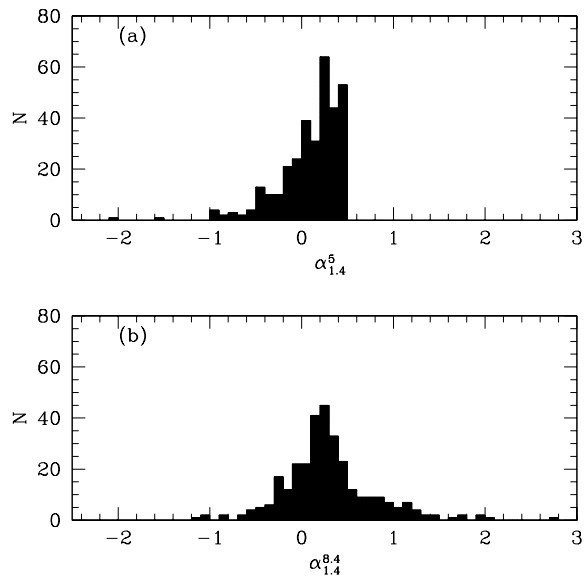,height=8cm,width=8cm}

 \caption{Distribution of the radio slopes between 1.4~GHz and 4.8~GHz
(a) and between 1.4~GHz and 8.4~GHz (b). Both panels do not include
lower limits}
\label{slopes2}
\end{figure}

1) Objects showing a complex morphology, with 2 or more components on scales
of 2$\arcmin$-4$\arcmin$: 15 sources. 

2) Objects not resolved in many components (45 sources). Some  of these 
are extended at the NVSS resolution (S$_{int}$/S$_{peak}>1.1)$ and others
are compact (S$_{int}$/S$_{peak}\le1.1$).

In case 1 the GB6 flux is expected to be the sum of most if not all the
components observed in the radio maps. The 1.4~GHz flux, instead, is the
sum of the fluxes within 70$\arcsec$ from the GB6 position, as
previously explained. Hence, for sources with an extension larger than
70$\arcsec$ the most external components are not included in the
computation of the 1.4~GHz flux. The resulting $\alpha_{1.4}^{4.8}$ is
thus flatter than the $\alpha_{1.4}^{8.4}$ computed with the total
fluxes.  For this reason the spectral index between 1.4 and 4.8~GHz was
recomputed in these cases by adding all the components resolved in the
NVSS maps and associated with the source itself. All sources turned out to
have $\alpha_{1.4}^{4.8}\ge$0.5 and they have thus been flagged as
``steep'' (an ``s'' in the flag column of Table~1; the corrected values
of $\alpha_{1.4}^{8.4}$ are reported between parenthesis).

In case 2, the most probable reason for the observed steep
$\alpha_{1.4}^{8.4}$ is that some flux is missed at 8.4~GHz. In fact,
the VLA (A array) observations are expected to miss flux on scales
larger than 6$\arcsec$ and, thus, well within the beamsize of the NVSS
observations.  This means that these objects are probably not strongly
core dominated sources. The extreme example is the planetary nebula
NGC~6543 (the object not detected at 8.4~GHz, GB6J175833+663801) which
appears with $\alpha_{1.4}^{8.4} >$3. This source is very strong at
1.4~GHz (772~mJy) and at 4.8~GHz (875~mJy) but when observed at 8.4~GHz
with a beam size of 0.24$\arcsec$, it disappears.

For reasons of completeness, the NVSS maps of the entire sample were
inspected in order to identify all possible cases of sources that fall
in case (1) studied above.  This analysis revealed another 7 sources
that were also flagged as ``steep'' (flag=s) in the last column of
Table~1. (The NVSS maps for the 22 steep sources are shown in
Figure~\ref{maps}.)  It is interesting to note that the percentage of
steep spectrum sources found in this sample corresponds exactly to that
which is expected in the complete CLASS catalogue (Meyers et al. 2001).

In summary, the presence of a ``tail'' of some steep (between 1.4 and
8.4~GHz) sources is mainly due to the fact that at 8.4~GHz some flux is
missed in objects which are not very compact.

\begin{figure*}
  \centerline{ \psfig{figure=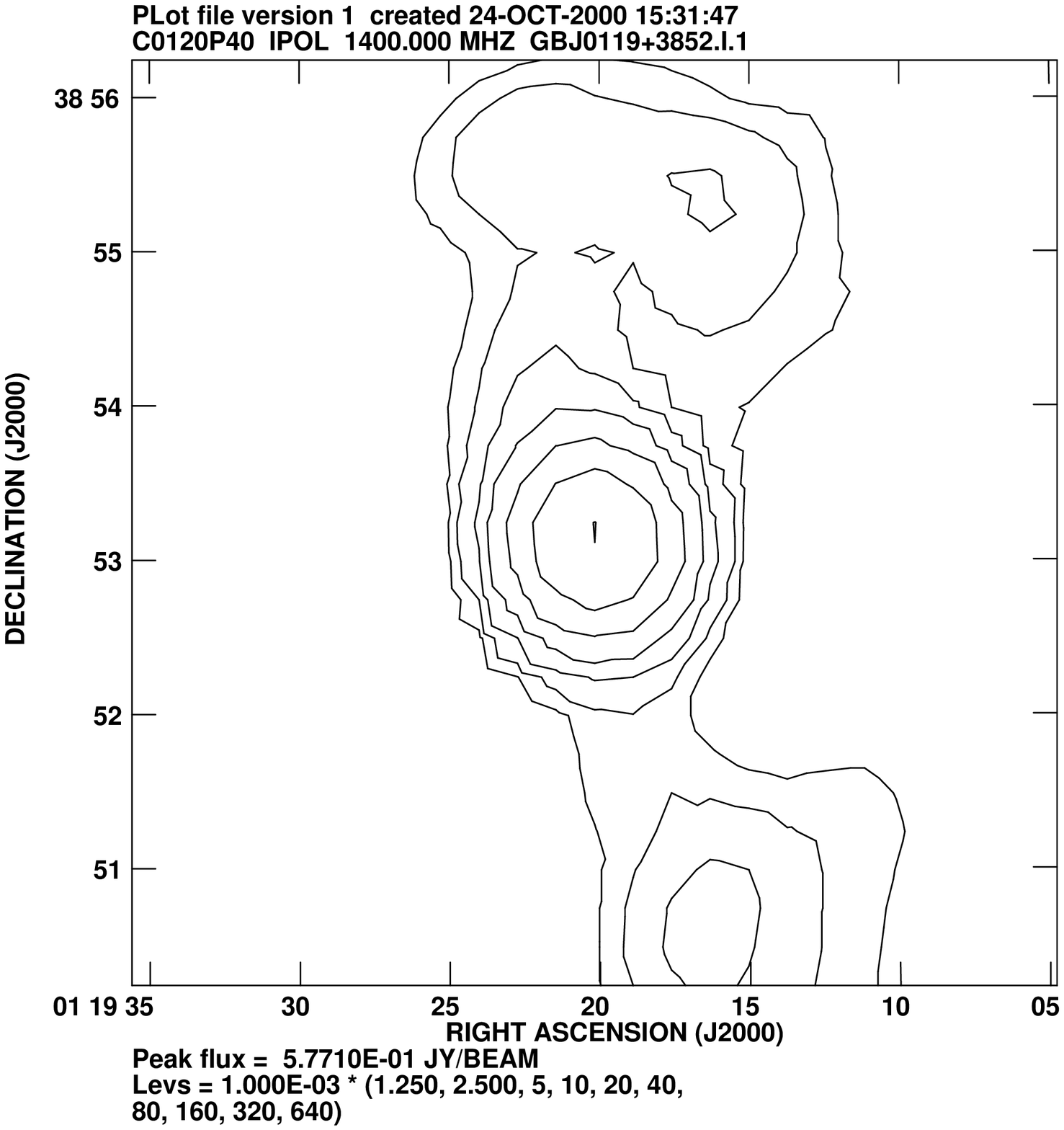,height=5cm,width=5cm}
    \psfig{figure=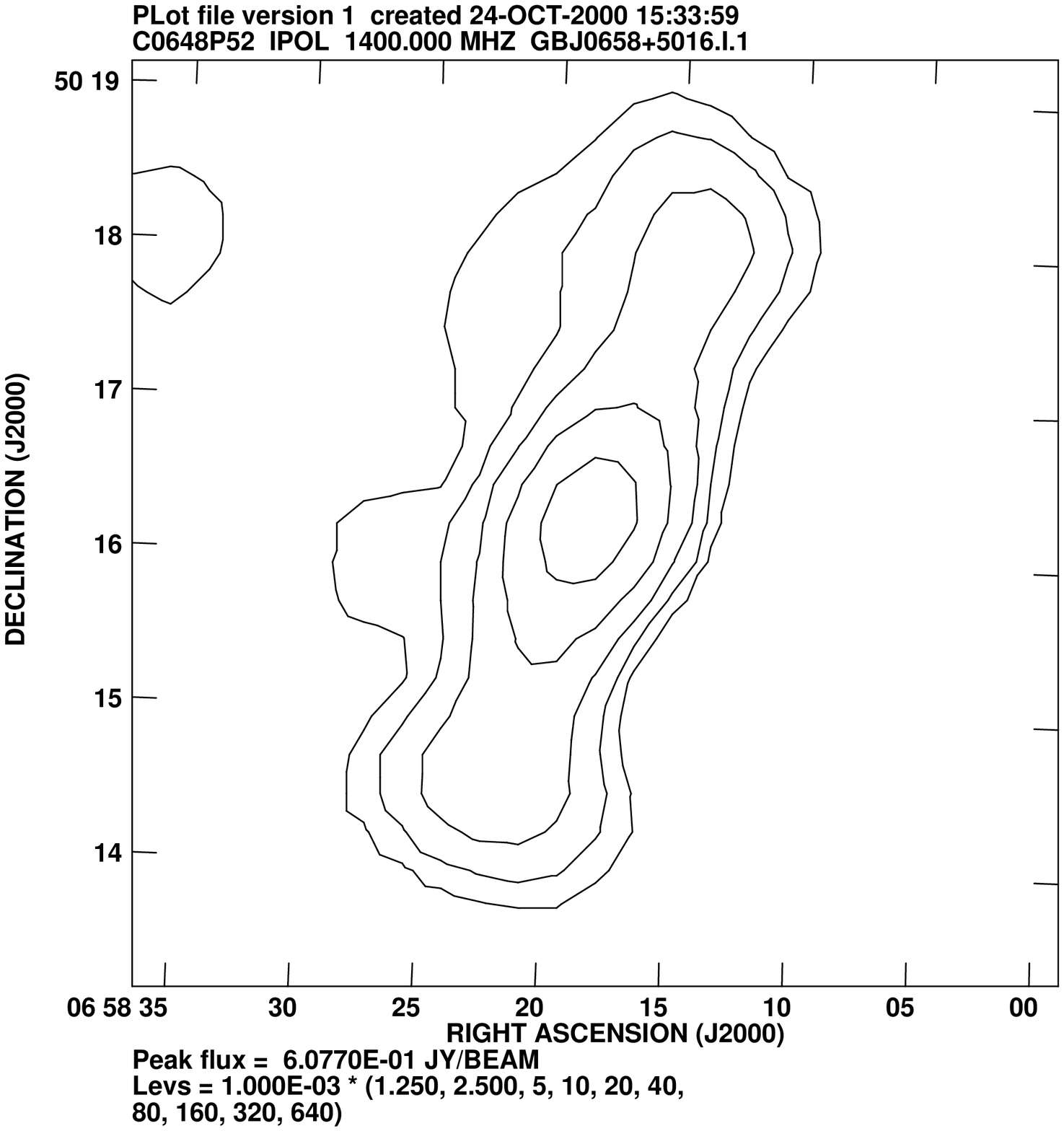,height=5cm,width=5cm}
    \psfig{figure=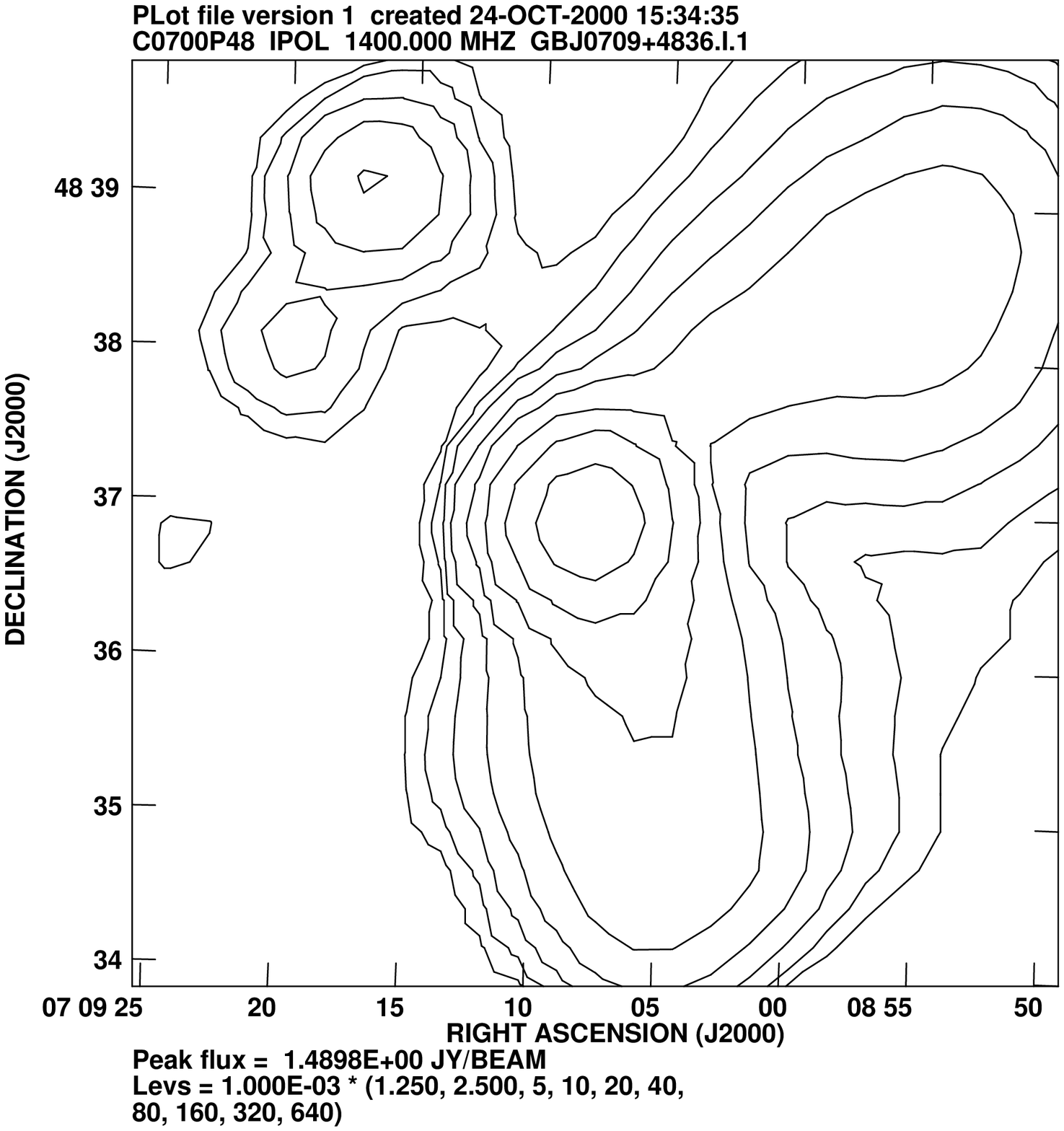,height=5cm,width=5cm} }
  \centerline{ \psfig{figure=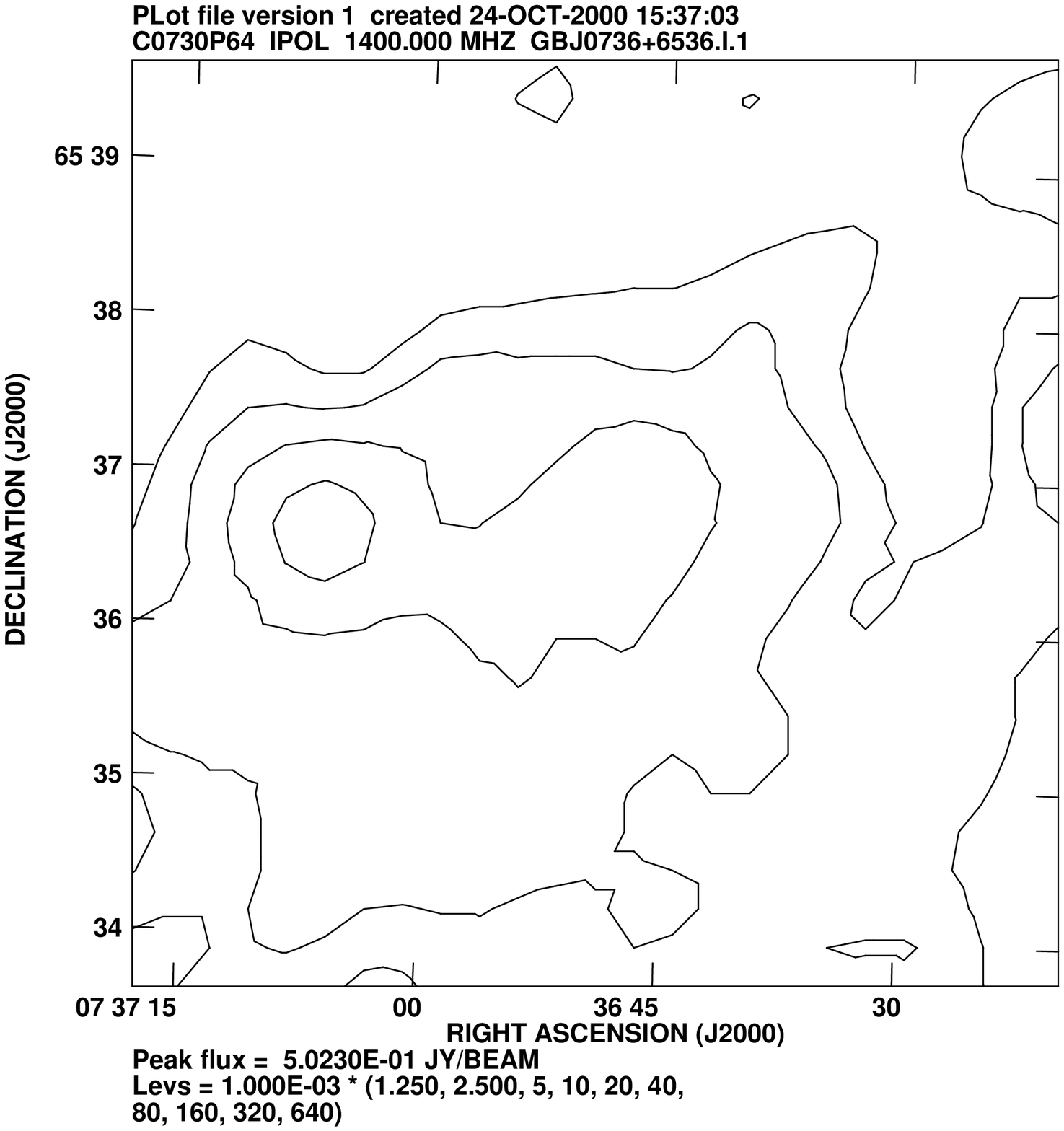,height=5cm,width=5cm}
    \psfig{figure=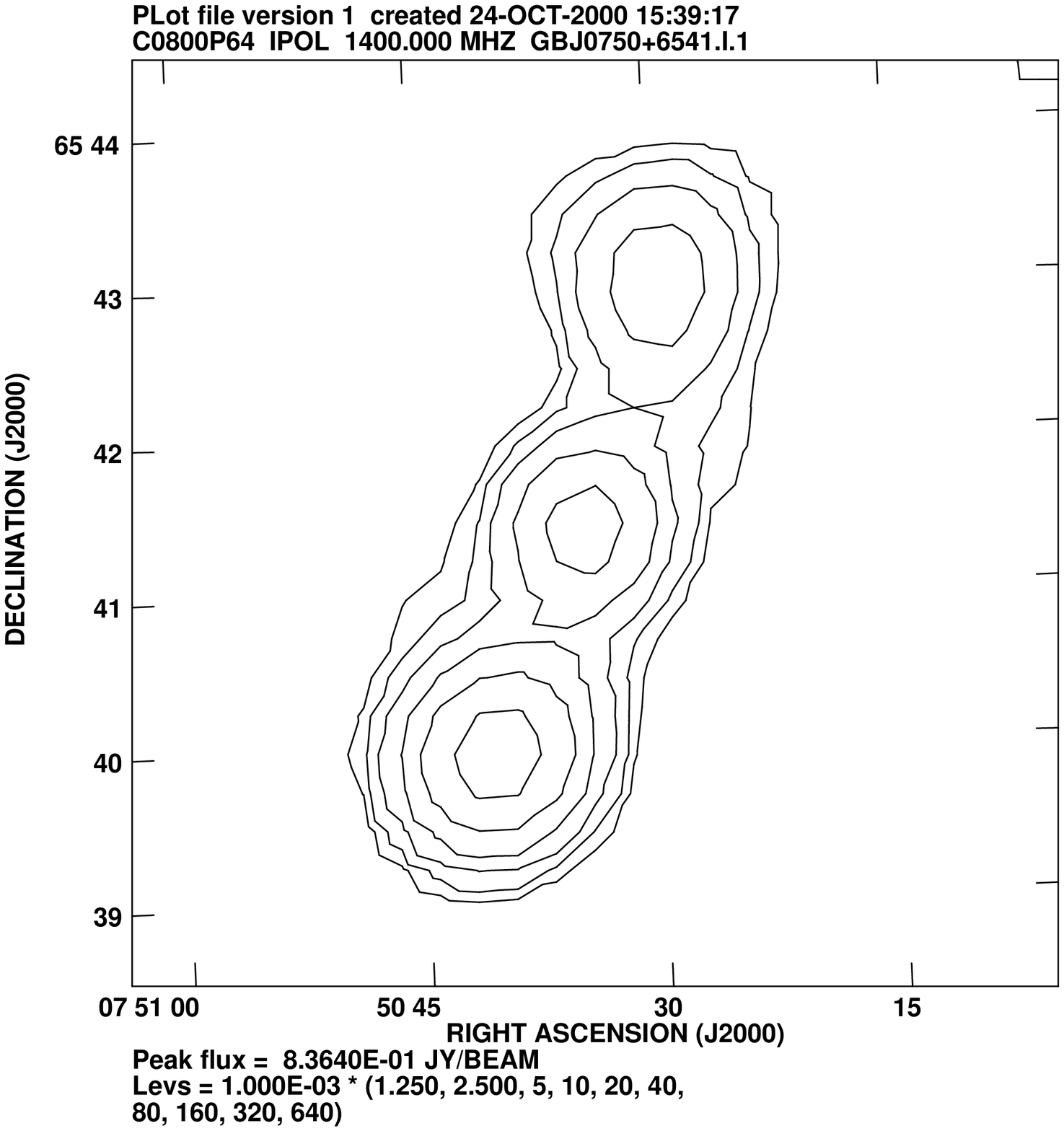,height=5cm,width=5cm}
    \psfig{figure=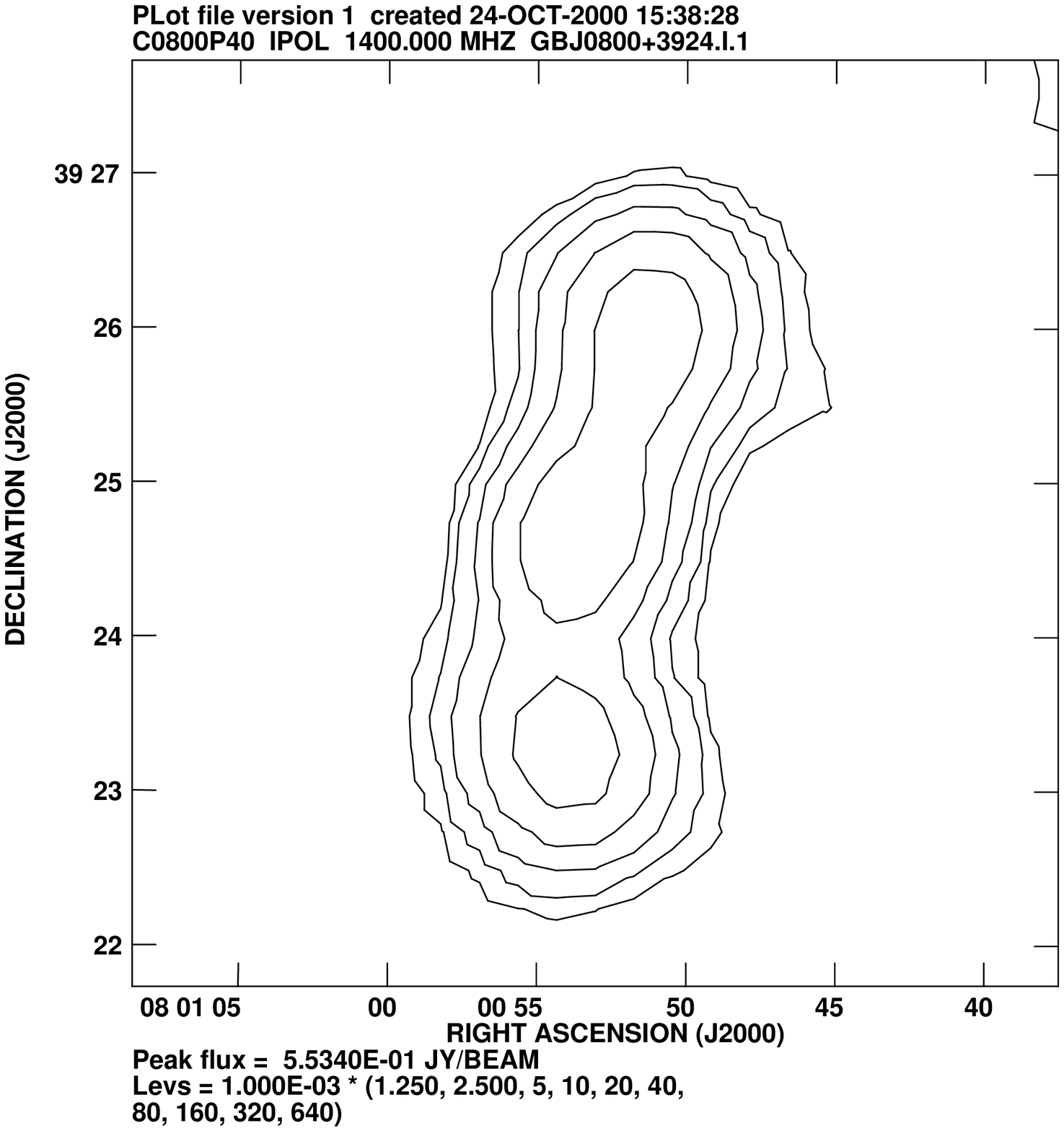,height=5cm,width=5cm} }

\centerline{
\psfig{figure=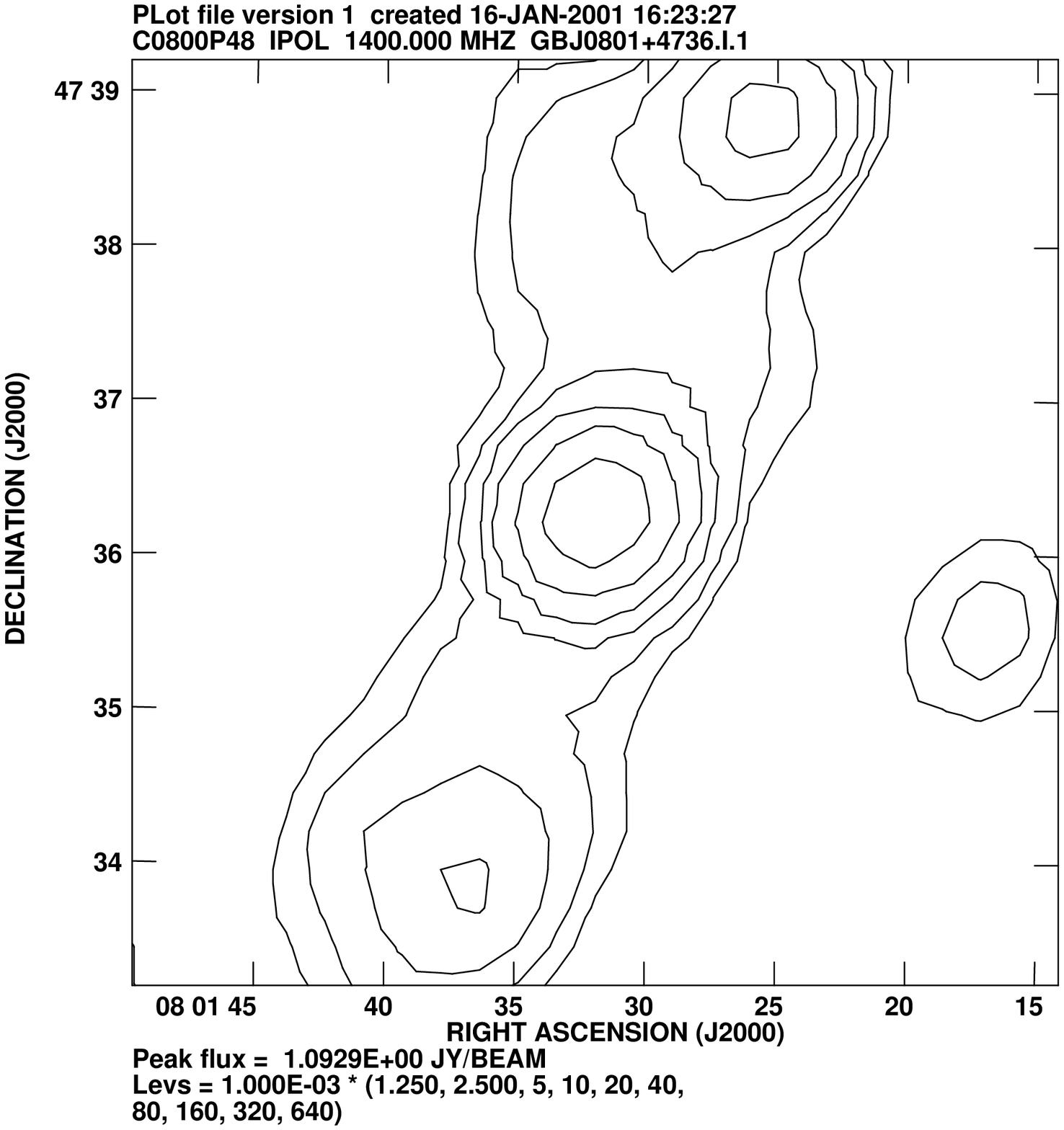,height=5cm,width=5cm}
\psfig{figure=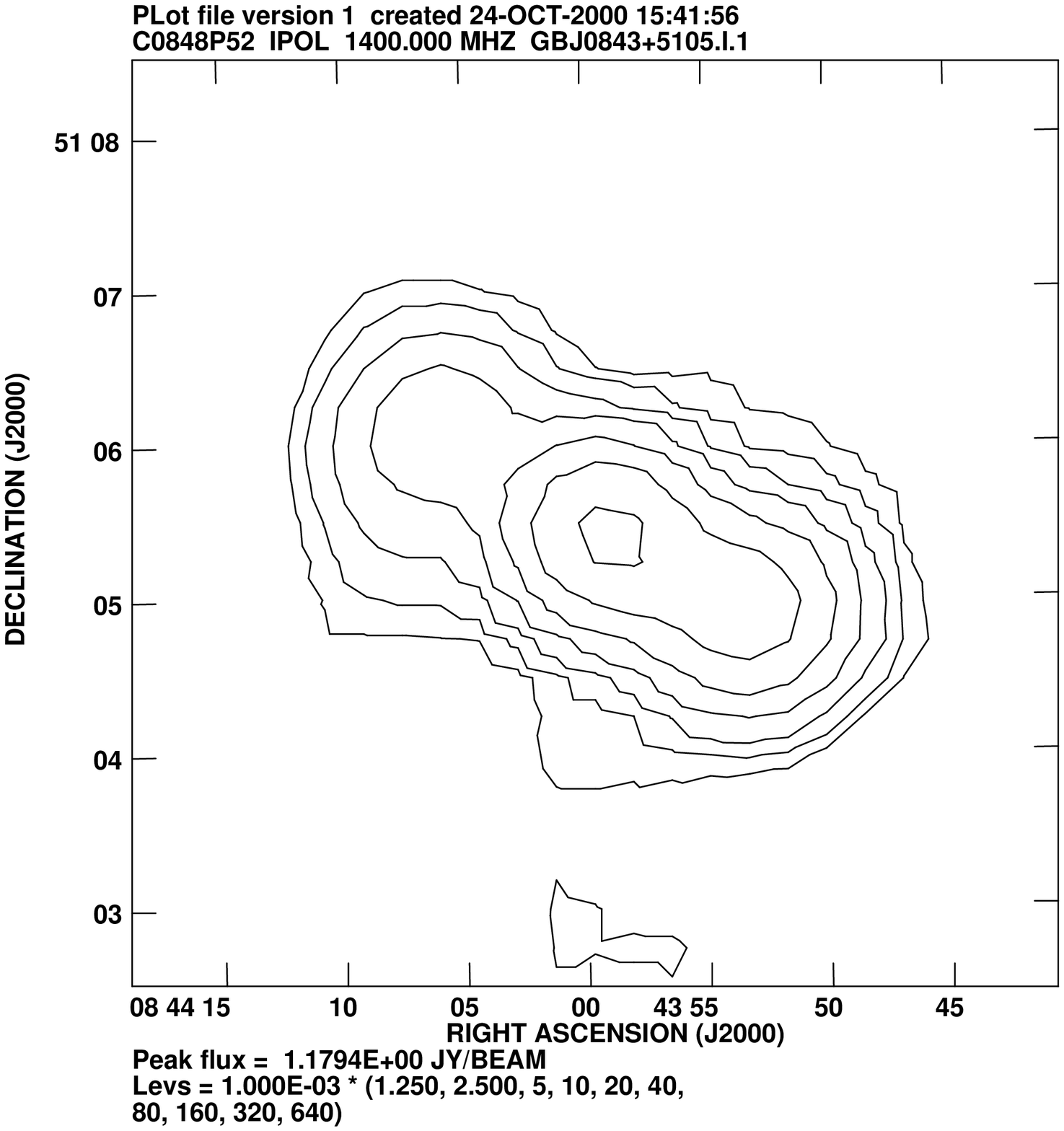,height=5cm,width=5cm}
\psfig{figure=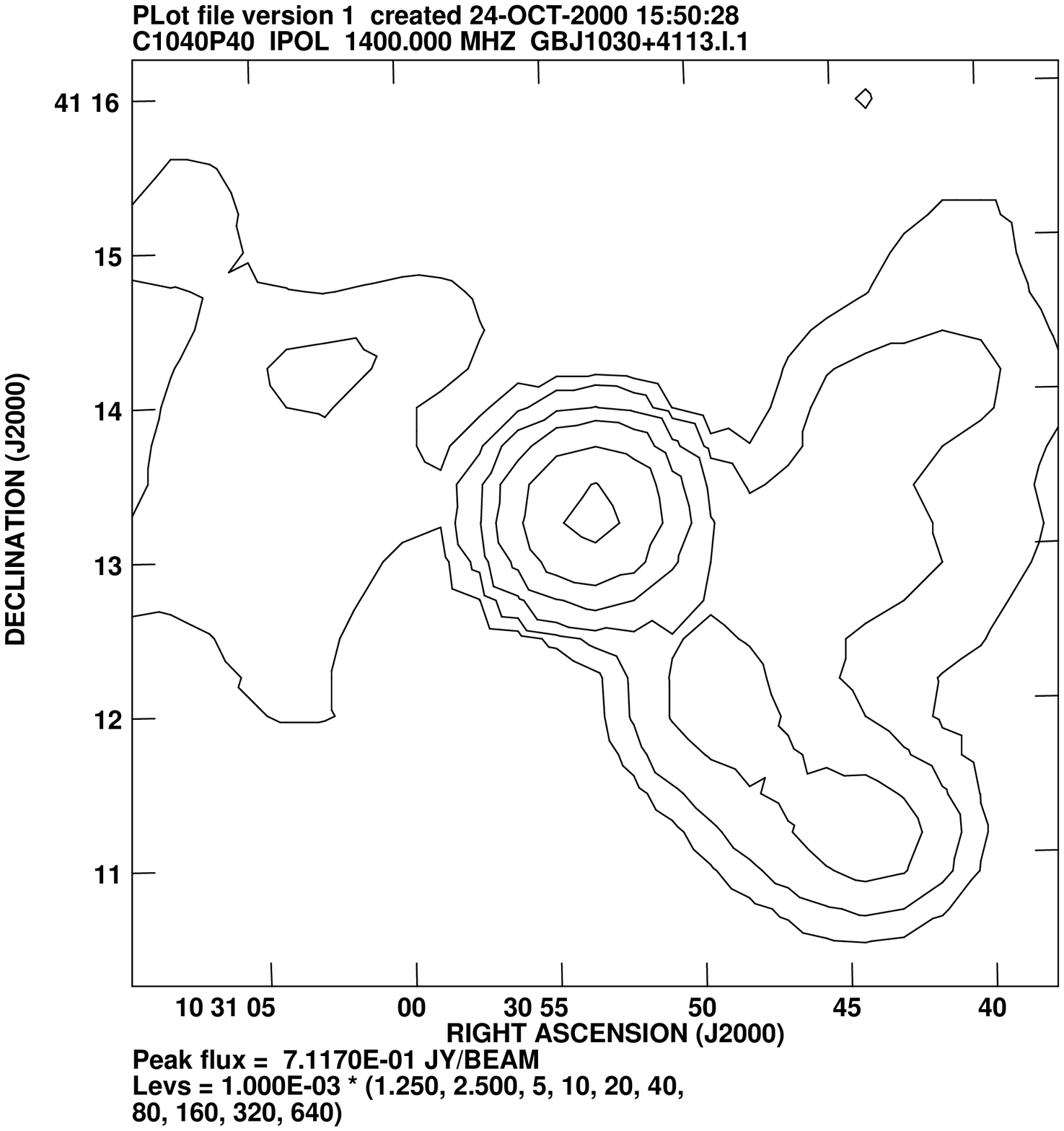,height=5cm,width=5cm}
}

\centerline{
\psfig{figure=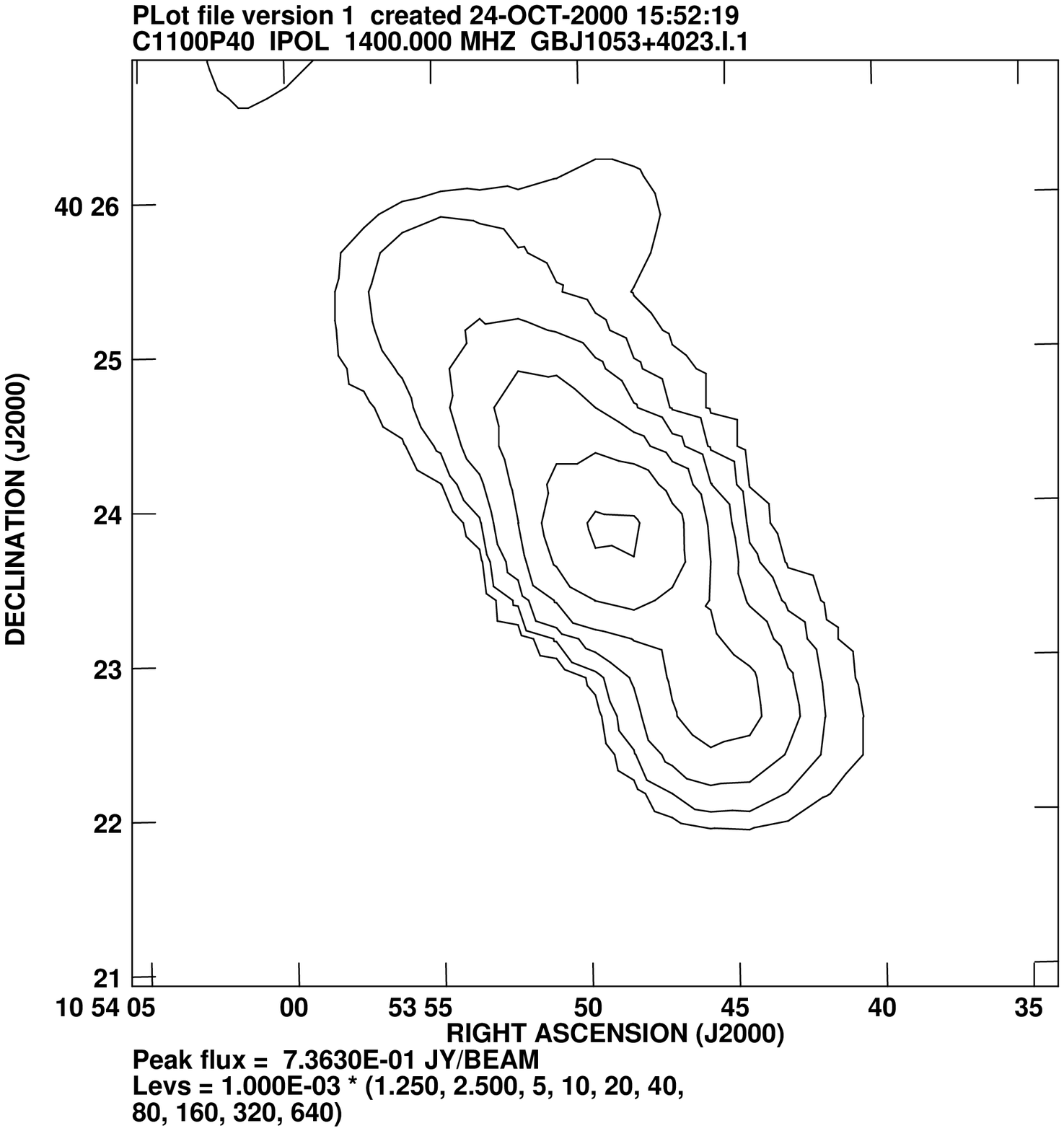,height=5cm,width=5cm}
\psfig{figure=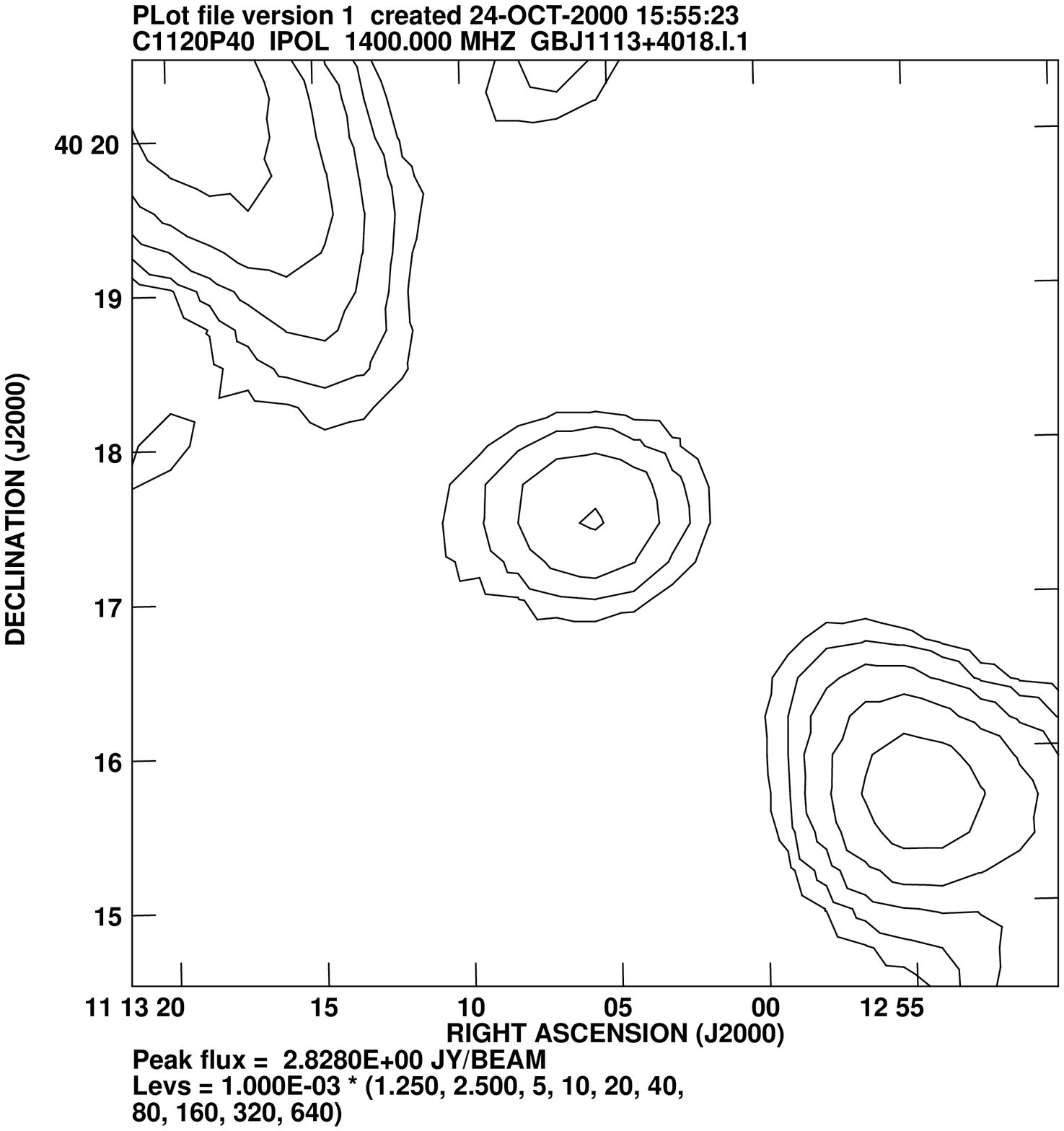,height=5cm,width=5cm}
\psfig{figure=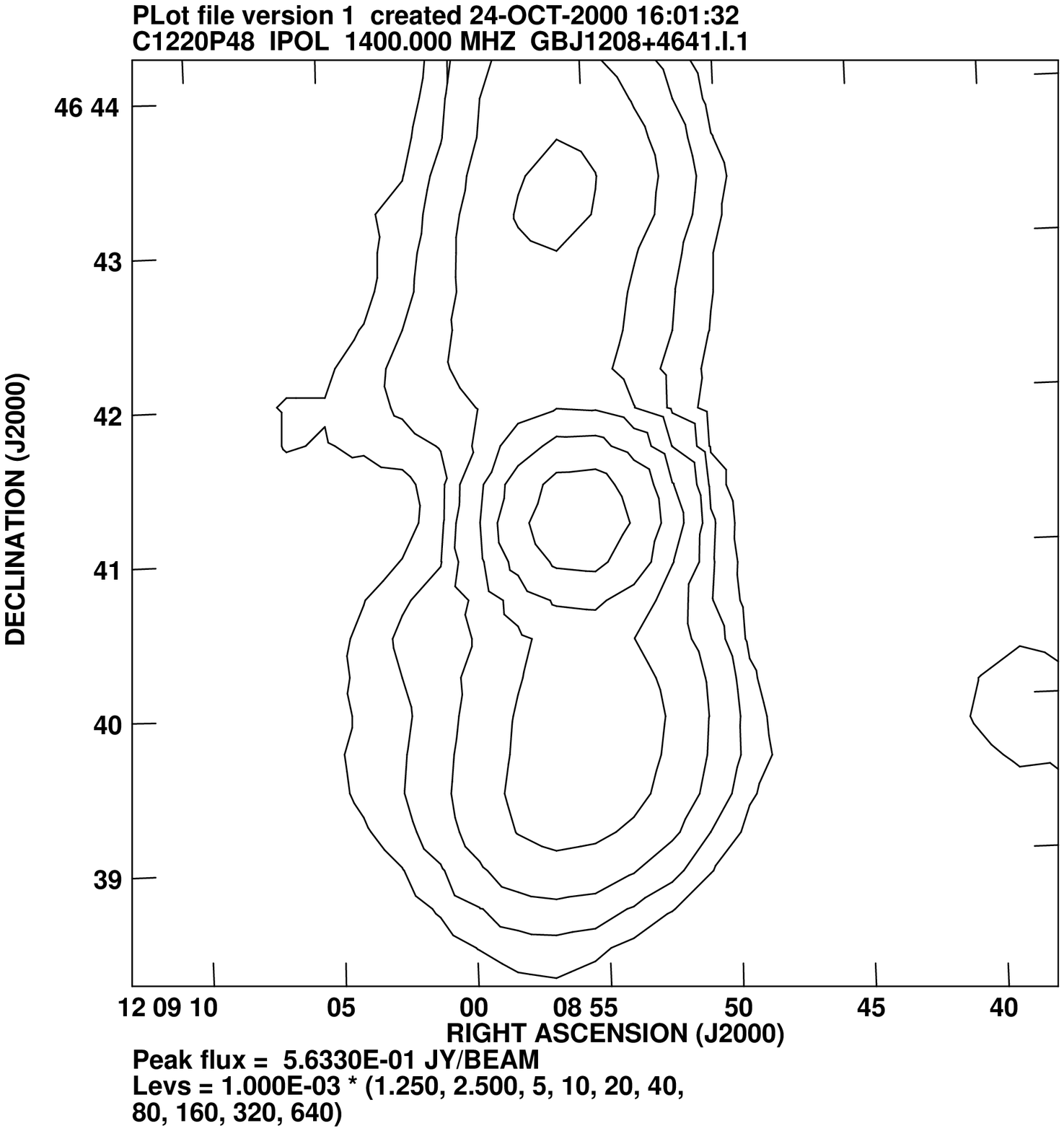,height=5cm,width=5cm}
}

\caption{The NVSS maps of the lobe-dominated steep spectrum 
sources discovered in the CLASS sample}
\label{maps}
\end{figure*}

 \begin{figure*}
\centerline{
\psfig{figure=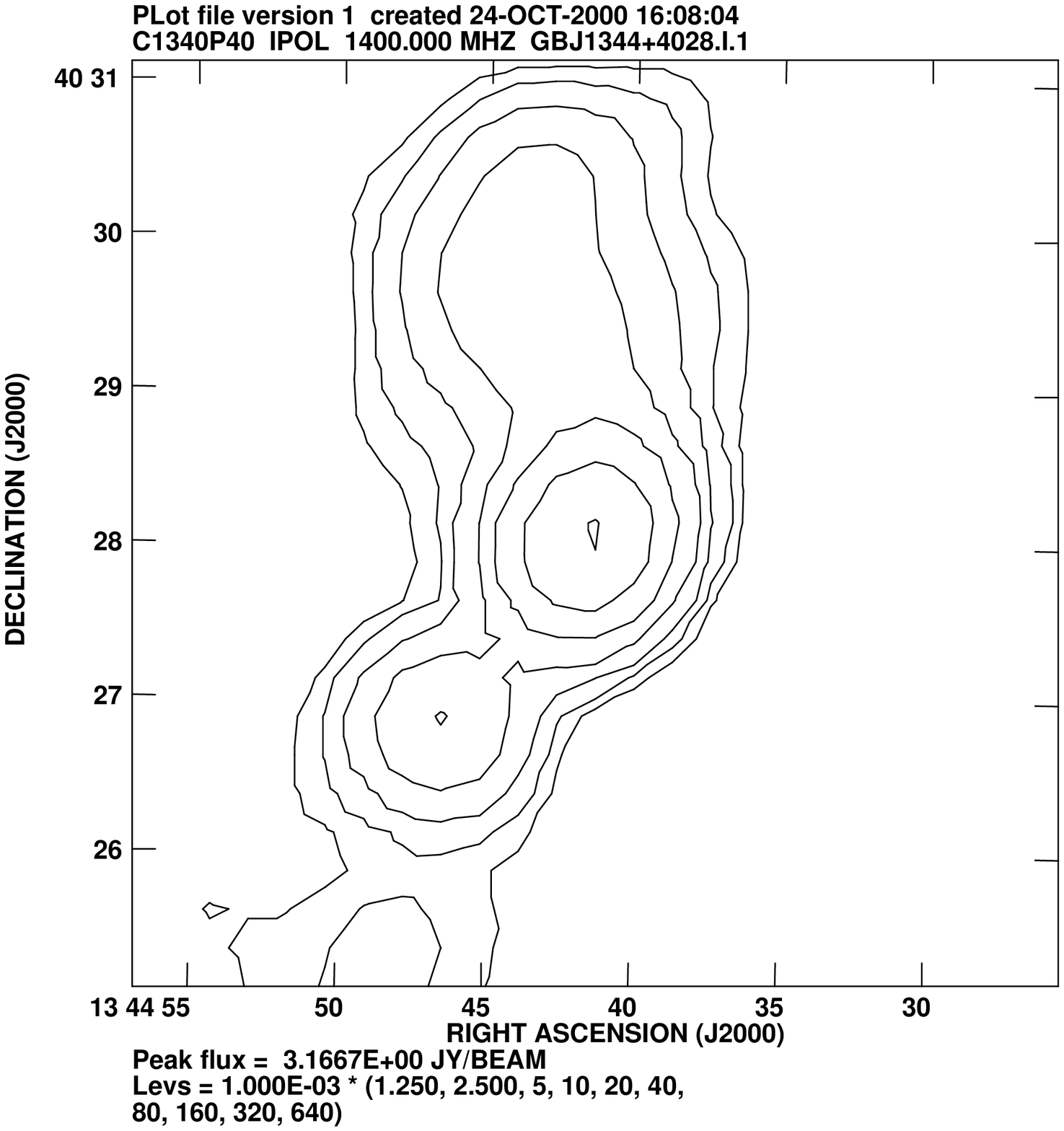,height=5cm,width=5cm}
\psfig{figure=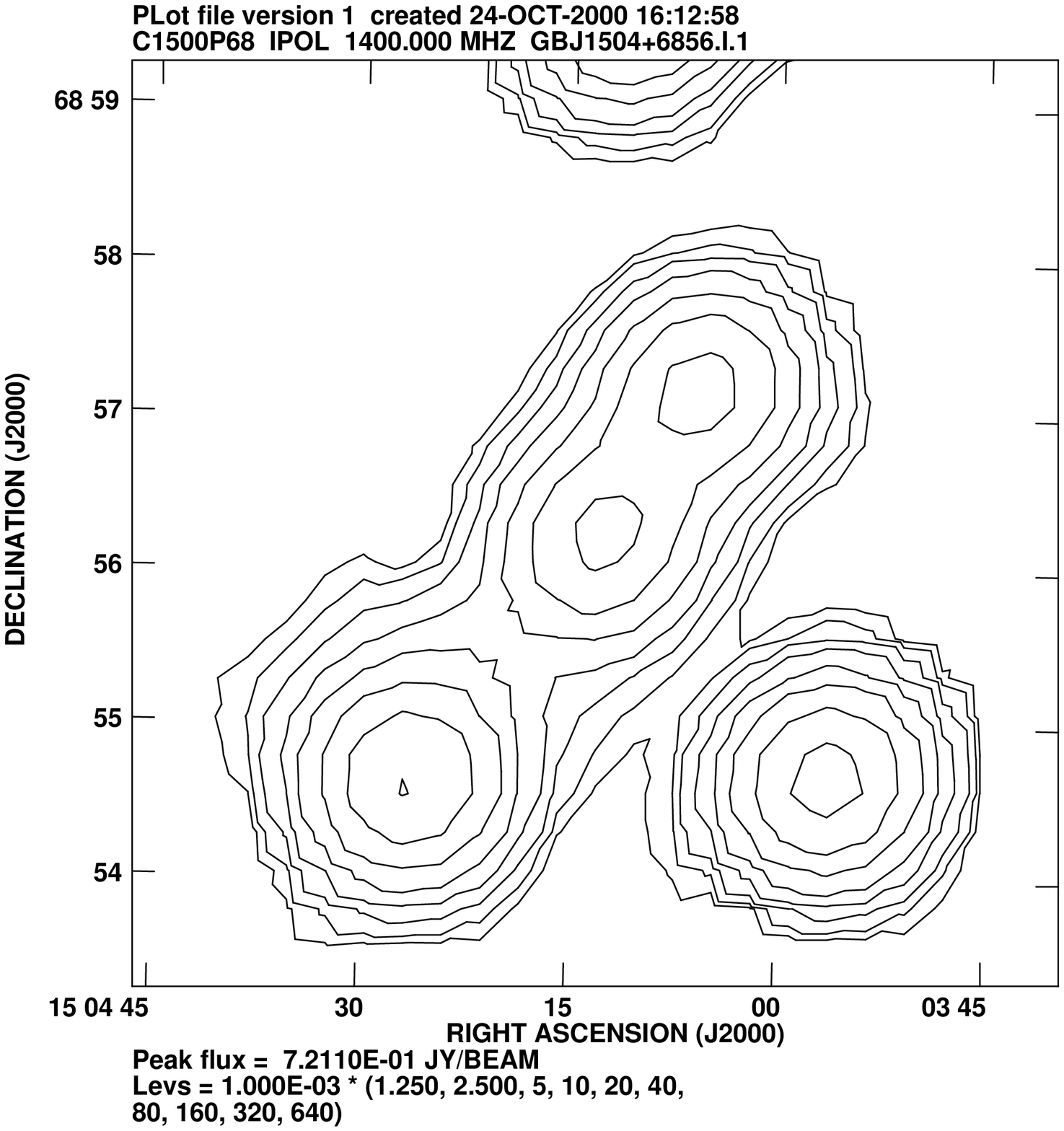,height=5cm,width=5cm}
\psfig{figure=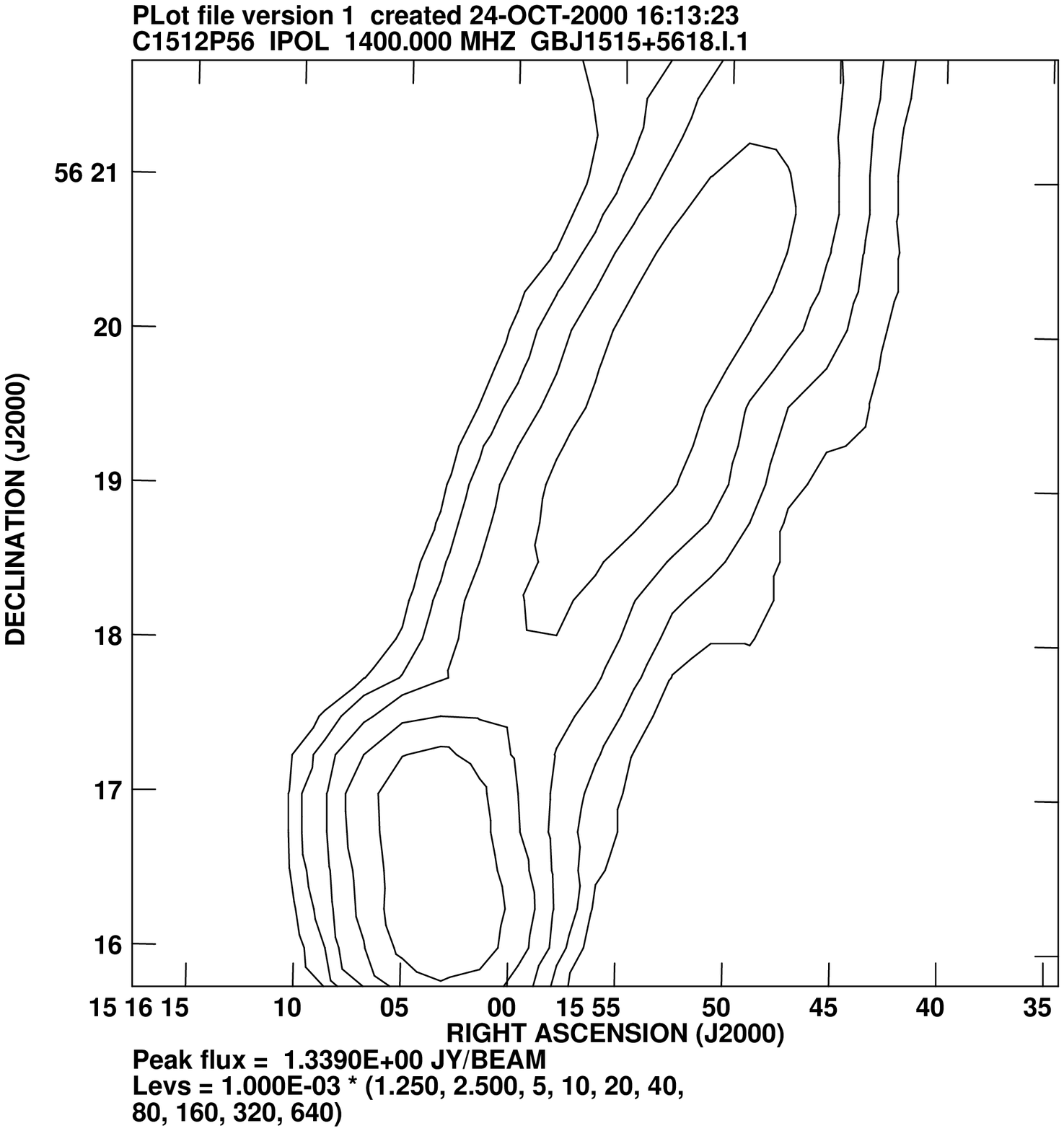,height=5cm,width=5cm}
}
\centerline{
\psfig{figure=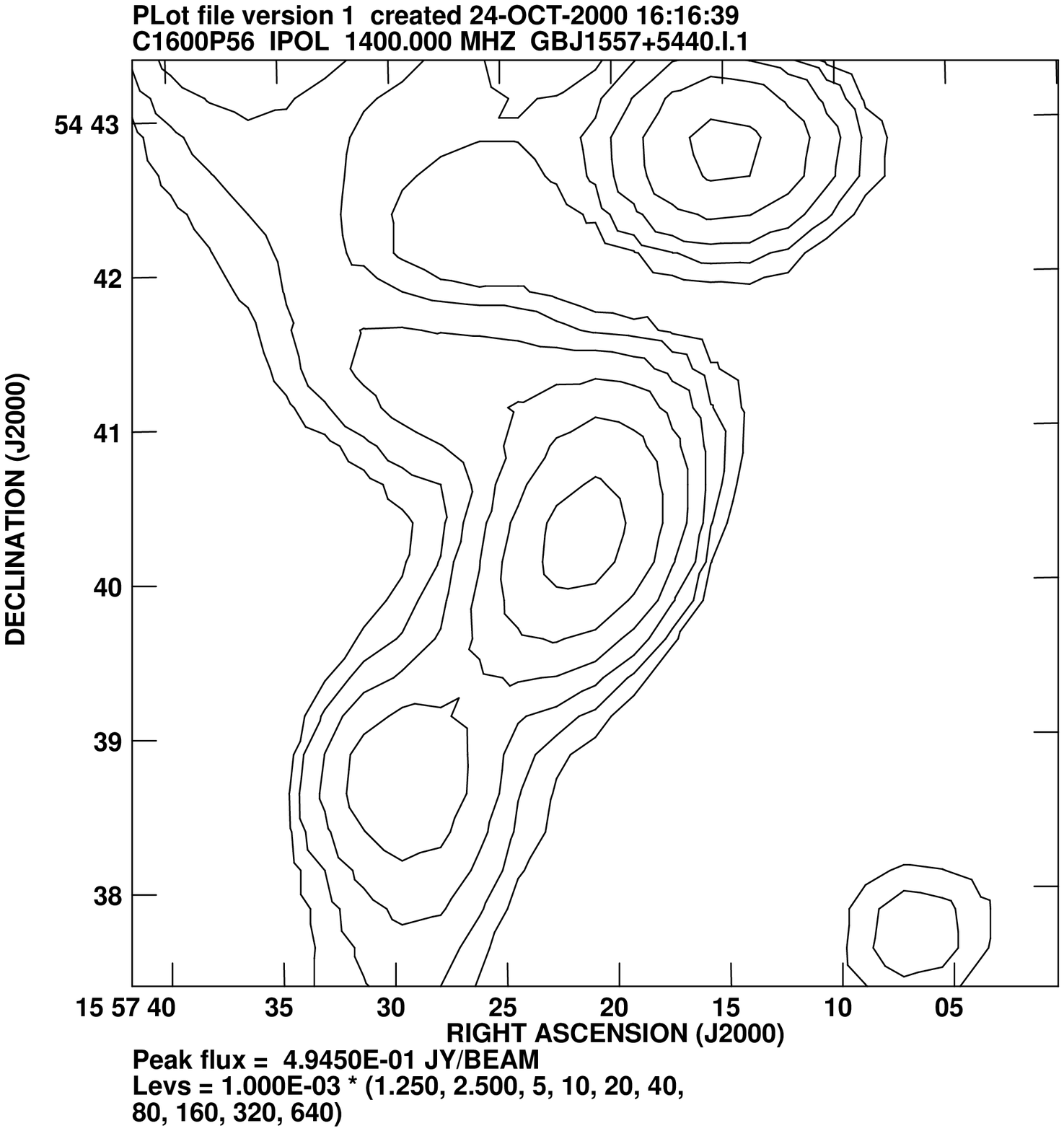,height=5cm,width=5cm}
\psfig{figure=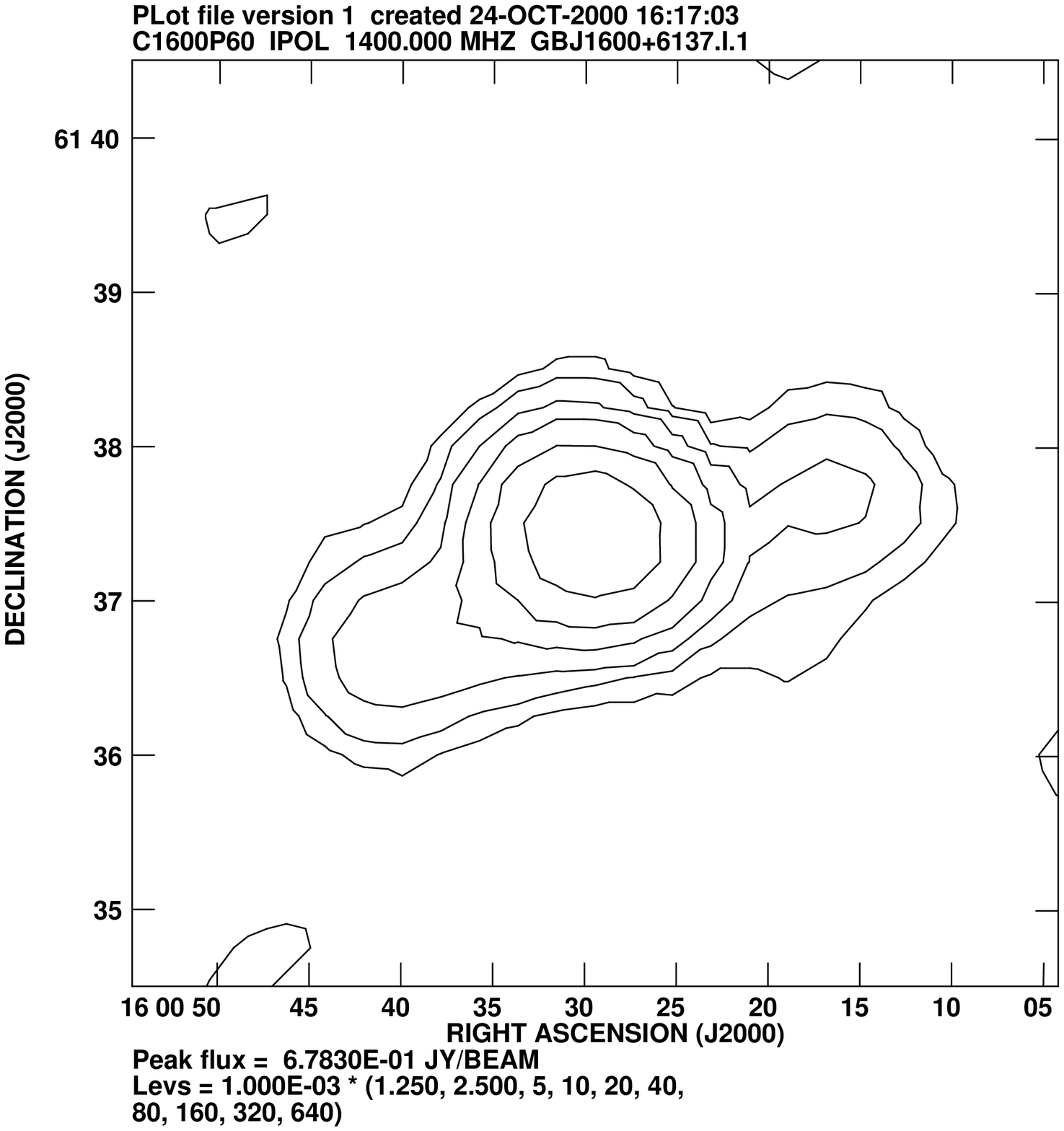,height=5cm,width=5cm}
\psfig{figure=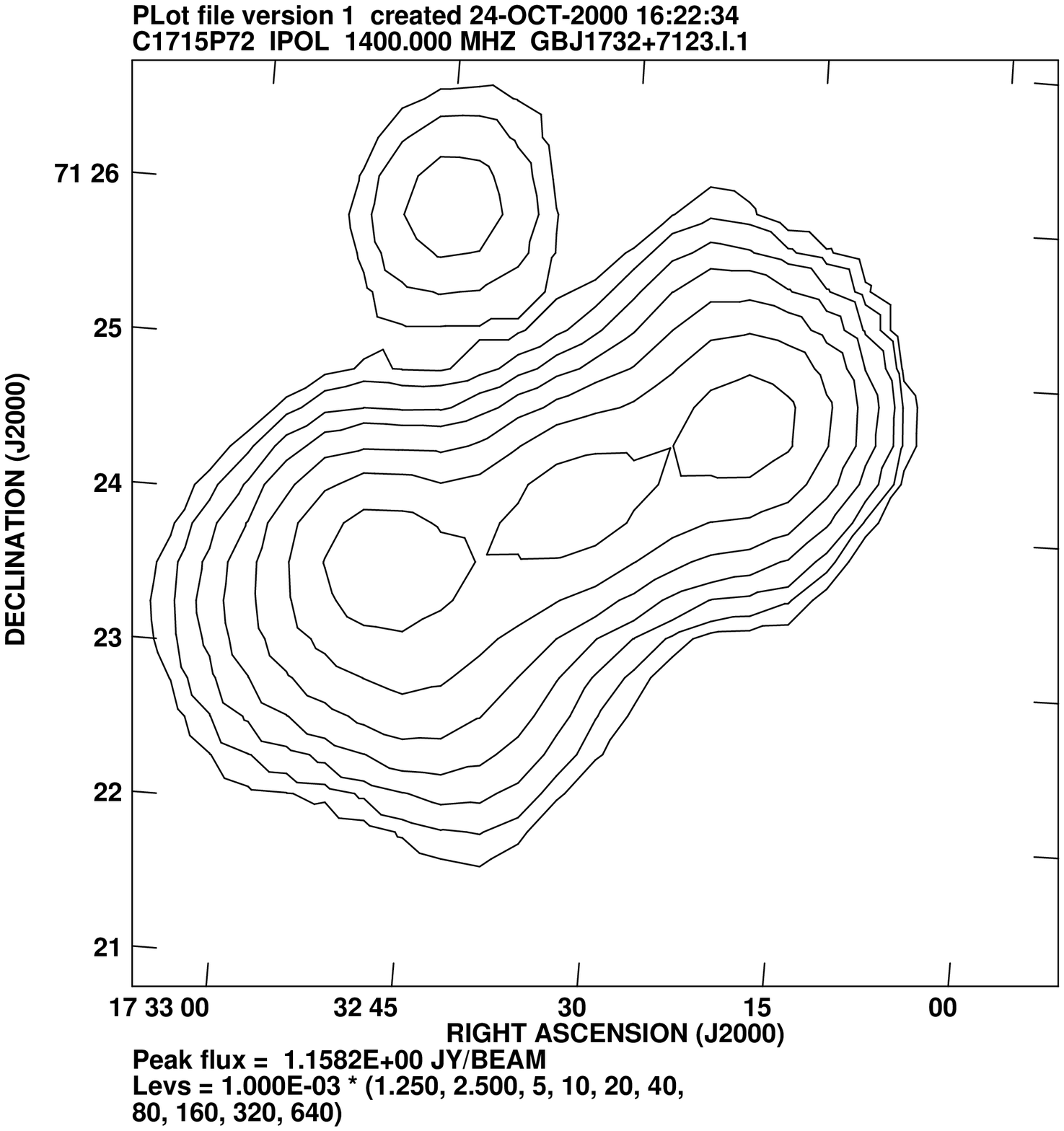,height=5cm,width=5cm}
}

\centerline{
\psfig{figure=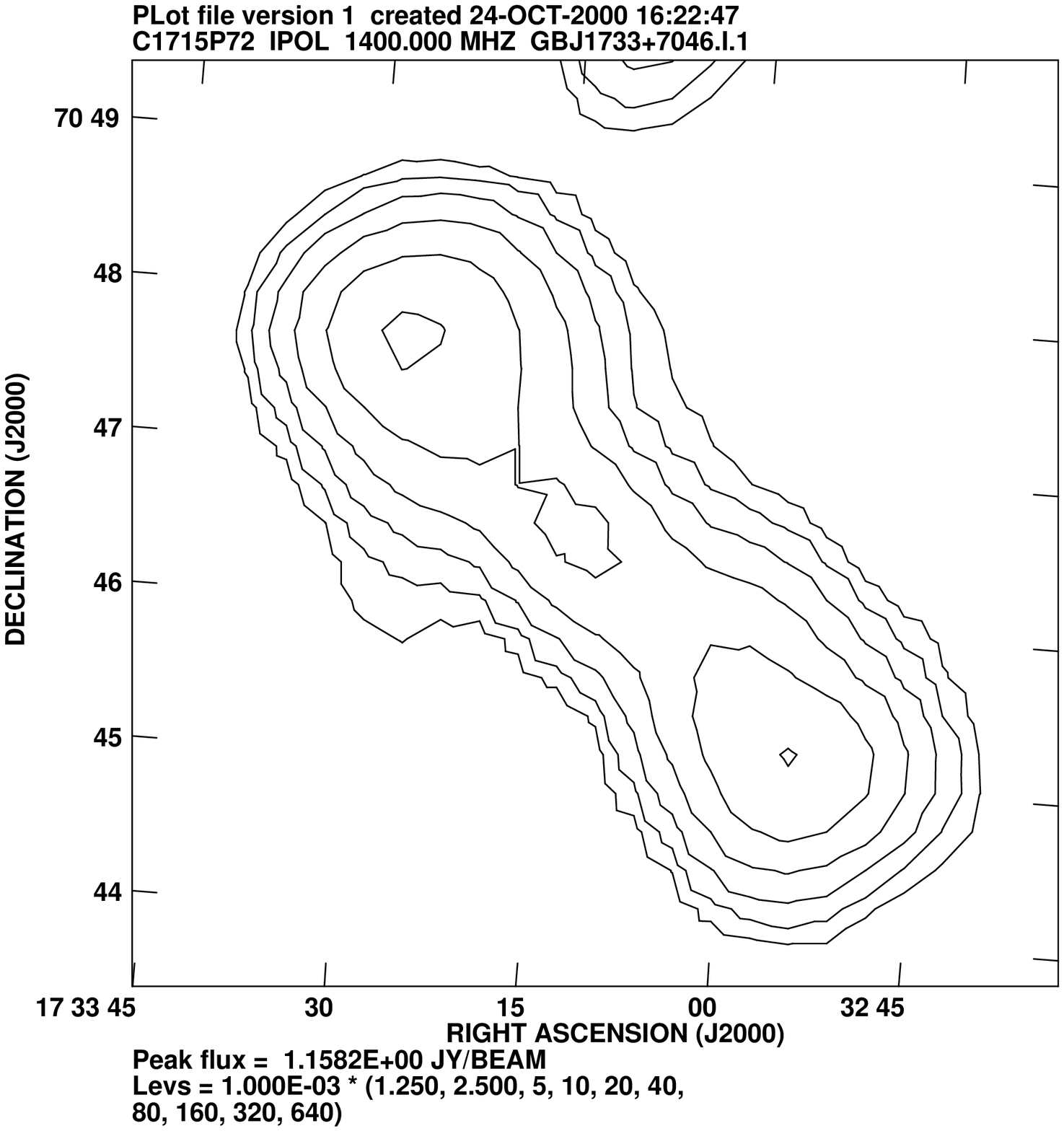,height=5cm,width=5cm}
\psfig{figure=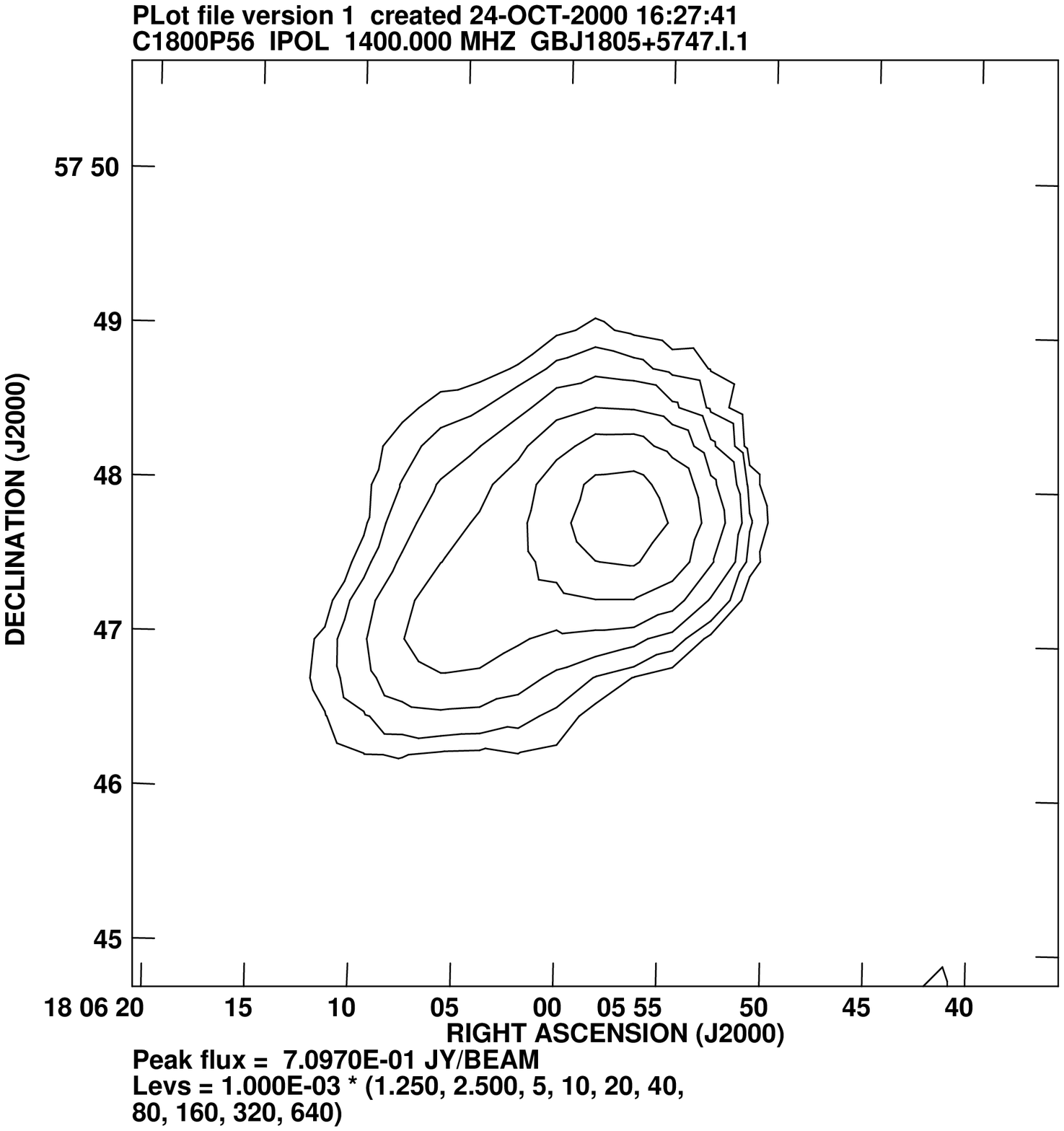,height=5cm,width=5cm}
\psfig{figure=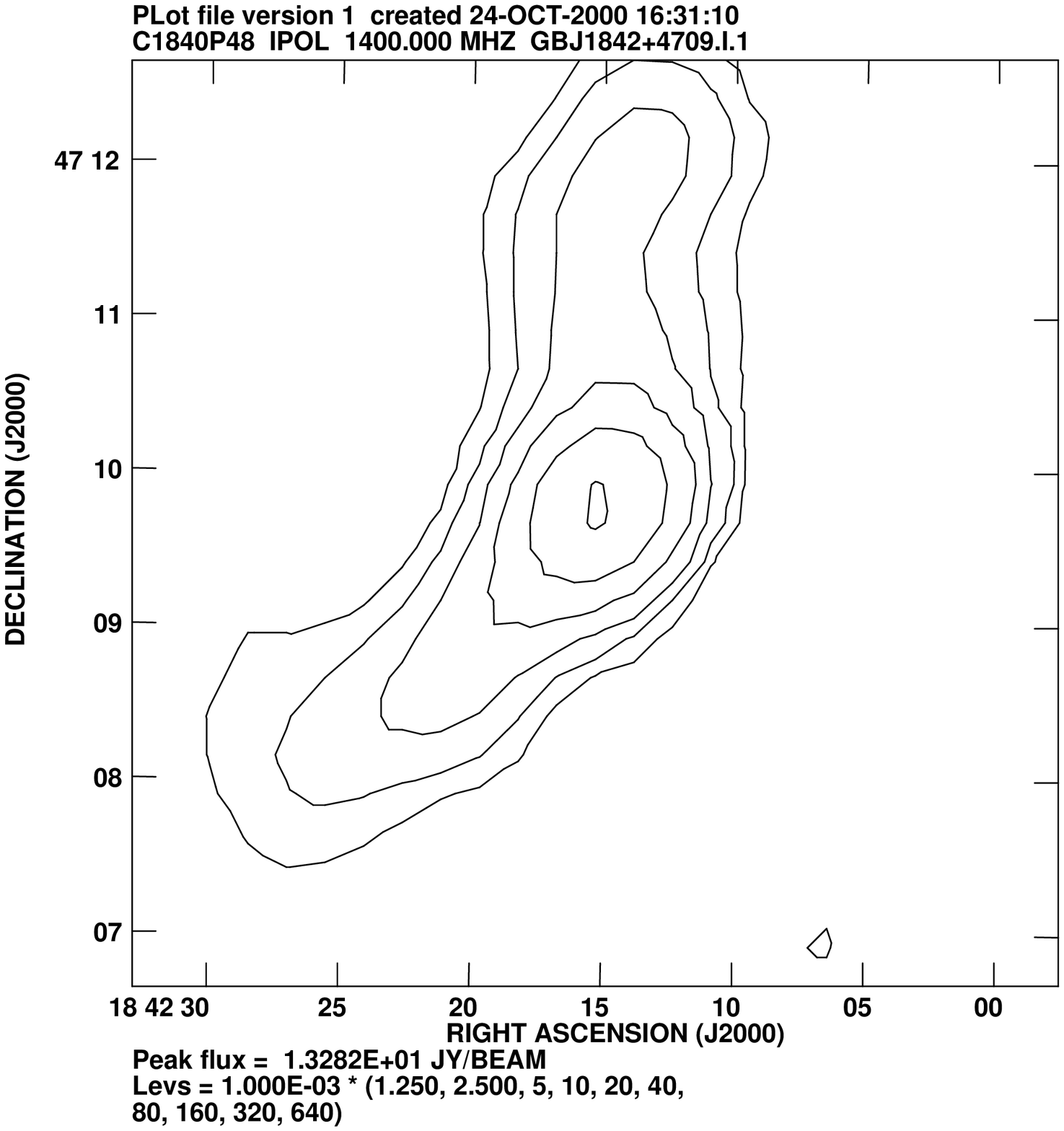,height=5cm,width=5cm}
}

\centerline{
\psfig{figure=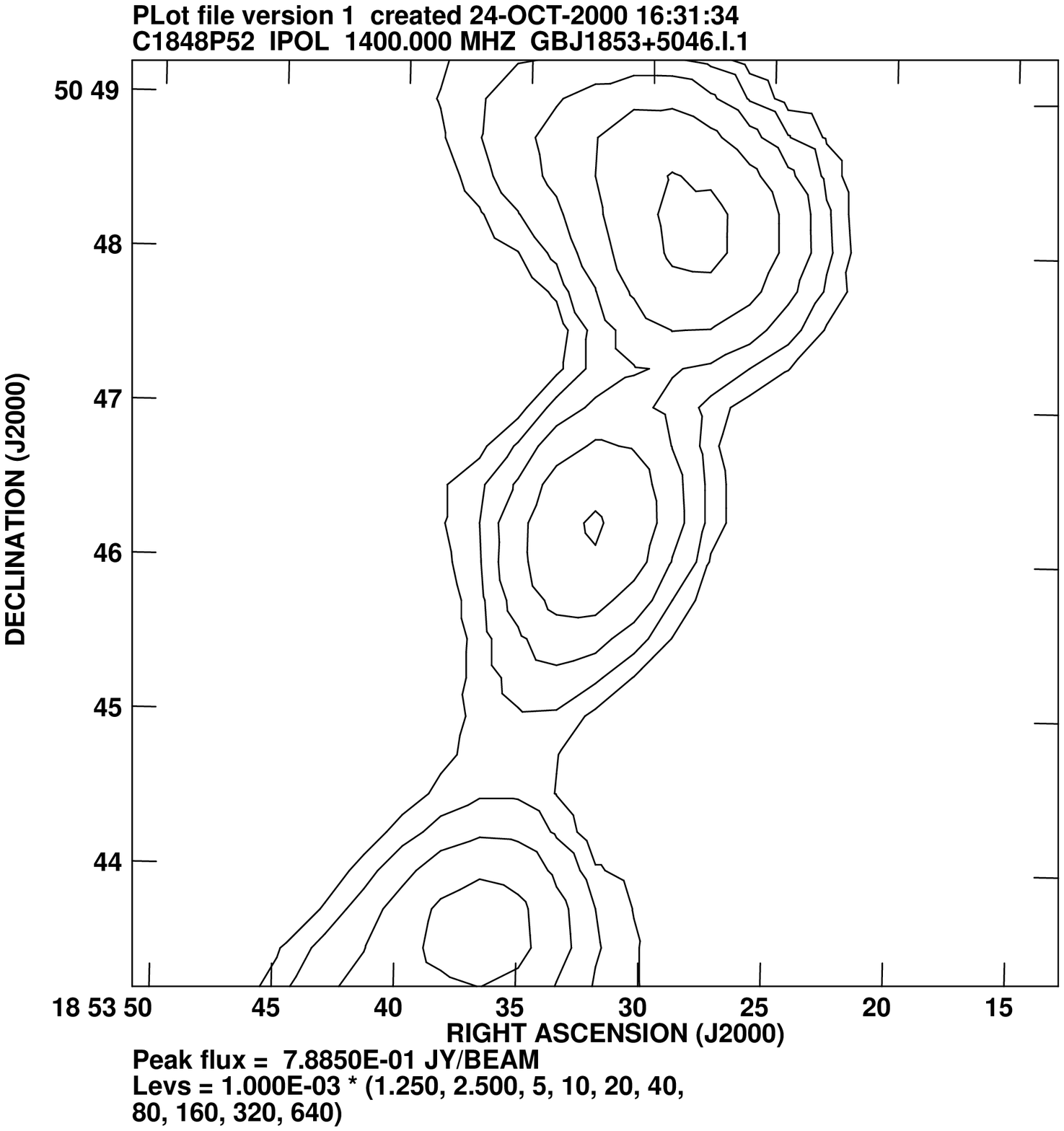,height=5cm,width=5cm}
}
\contcaption{}
\end{figure*}

 \begin{figure}
\psfig{figure=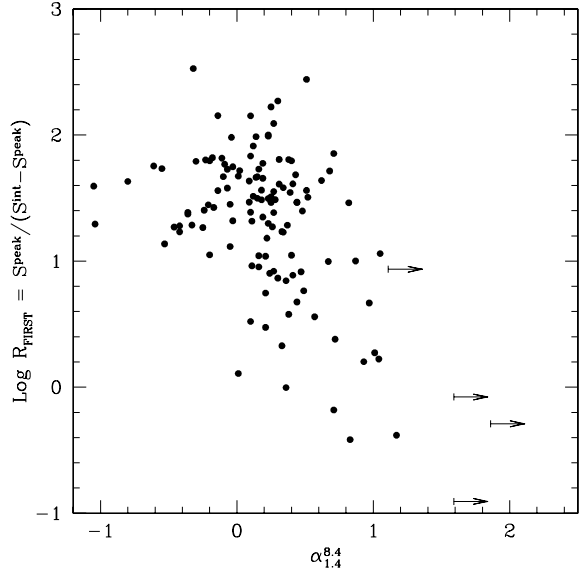,height=8cm,width=8cm}

 \caption{The radio ``compactness'' parameter (core/extended flux)
defined with the FIRST fluxes assuming core flux=peak flux and extended
flux=integrated -- peak flux. The symbols are the same as in 
Figure~\ref{slopes3} }
\label{first}
\end{figure}

\subsection{The extended sources}

Since 144 objects fall in the region of sky covered by the FIRST
survey, it is possible to compute a reliable radio ``compactness''
parameter. The 16 objects which are resolved by FIRST in 2 components
were excluded from the analysis.
  
Assuming:

\[ S_{core} = S^{FIRST}_{peak} \]

and 

\[ S_{extended} = S^{FIRST}_{total}-S^{FIRST}_{peak} \] 

the following parameter {\it R} is defined:

\[R = S_{core}/S_{extended} \]

In Figure~\ref{first} the computed R parameter versus
$\alpha_{1.4}^{8.4}$ is presented. Despite the large scatter observed in
this plot, it is possible to infer the following conclusions: (1) the
objects with a very flat (inverted) spectrum are all strongly
core-dominated with R$\geq$10; (2) the objects with steep spectrum
($\alpha_{1.4}^{8.4}\geq$0.5) cover a wide range of R parameter ranging
from from 0.1 up to 60 although the majority show R$\geq$1.  The
extended sources are probably steep because at 8.4~GHz the VLA
observations are missing a fraction of the extended flux (see discussion
in the previous section). Nevertheless, the majority of sources with a
steep $\alpha_{1.4}^{8.4}$ are compact in the FIRST maps. For these,
variability could be another possible explanation for the observed
difference between $\alpha_{1.4}^{8.4}$ and $\alpha_{1.4}^{4.8}$.

\section{Conclusion}

A new deep sample of flat radio spectrum sources based on CLASS data
has been presented. This sample has been designed to select and study
the low-luminosity tail of the {\it blazar} class of AGN which, under
the current unification model, should contain
mainly optically featureless objects (BL Lacs).  The goal is to test
this hypothesis by systematically studying the optical properties of a
well defined and complete local sample of blazars. To this end, only
the optically bright (R$\leq$17.5) CLASS sources have been considered.
The resulting sample contains 325 objects and it is the
natural extension of the 200~mJy sample by March\~a et al. (1996) down
to fainter fluxes (30~mJy at 4.8~GHz).

In this first paper the radio properties of these 325 objects have been
studied by using the available data at different frequencies and
resolutions. It has been concluded that the vast majority of the objects
selected in the CLASS blazar survey are compact (R$\geq$1) even at 
the 6$\arcsec$ beam of the FIRST survey with a large percentage 
($\sim$76\%) of objects with very high R ($\geq$10).
Furthermore, it has also been shown that the spectral
index between 1.4 and 4.8~GHz is well correlated with the index computed
between 1.4 and 8.4~GHz. Finally, only 22 objects show clear
lobe-dominated radio structures which owe their appearance in the sample
to the different resolution of the catalogues used for the
selection.

The optical follow-up of the entire sample is in progress and, so far,
more than 70\% of the objects have been spectroscopically classified.
The results from the optical identification will be presented and
discussed in a forthcoming paper (Caccianiga et al.; submitted to
MNRAS).

\section*{Acknowledgments}
This research was supported in part by the European Commission
Training and Mobility of Researchers program, research 
network contract ERBFMRX-CT96-0034 ``CERES''.

\end{document}